\documentclass[12pt]{JHEP3}
\usepackage{amsfonts}
\usepackage{epsfig}%
\usepackage{amsmath}%
\usepackage{amsfonts}%
\usepackage{amssymb}%
\usepackage{graphicx}

\usepackage[perpage,symbol*]{footmisc}

\def\SigmapN{\Sigma_{\pi N}}
\def\addcite#1{[???]}
\def\chisb{\raise0.2em\hbox{$\chi$}SB}
\def\chiSM{\raise0.17em\hbox{$\chi$}SM}

\def\gray{\special{ps: 0.4 setgray}}
\def\black{\special{ps: 0.0 setgray}}

\newcommand{\draft}{
\newcount\timecount
\newcount\hours \newcount\minutes  \newcount\temp \newcount\pmhours

\hours = \time
\divide\hours by 60
\temp = \hours
\multiply\temp by 60
\minutes = \time
\advance\minutes by -\temp
\def\hour{\the\hours}
\def\minute{\ifnum\minutes<10 0\the\minutes
            \else\the\minutes\fi}
\def\clock{
\ifnum\hours=0 12:\minute\ AM
\else\ifnum\hours<12 \hour:\minute\ AM
      \else\ifnum\hours=12 12:\minute\ PM
            \else\ifnum\hours>12
                 \pmhours=\hours
                 \advance\pmhours by -12
                 \the\pmhours:\minute\ PM
                 \fi
            \fi
      \fi
\fi
}
\def\fullclock{\hour:\minute}
\gray
\font\Hugett  =cmtt12 scaled\magstep2
{\Hugett \strut \kern-3em Draft: \today,\clock}
\black
} 
\catcode`\@=11 
\def\lsim{\mathrel{\mathpalette\@versim<}}
\def\gsim{\mathrel{\mathpalette\@versim>}}
\def\@versim#1#2{\vcenter{\offinterlineskip
        \ialign{$\m@th#1\hfil##\hfil$\crcr#2\crcr\sim\crcr } }}
\catcode`\@=12 

\def\lappeq{\lsim}

\def\nextline{\hfill\break}

\def\mycomm#1{\nextline\strut\kern-6em{\tt ====> \ #1}\nextline}
\def\mpcomm#1{\nextline\strut\kern-6em{\tt MP COMMENT => \ #1}\nextline}
\def\nextline{\hfill\break}

\newcommand{\beq}{\begin{equation}}
\newcommand{\eeq}{\end{equation}}
\newcommand{\bea}{\begin{eqnarray}}
\newcommand{\eea}{\end{eqnarray}}
\newcommand{\Seff}{S_{\kern-0.1em\hbox{\small \it eff}}}
\newcommand{\Leff}{{\cal L}_{\kern-0.1em\hbox{\small \it eff}}}
\newcommand{\SMB}{S_{\kern-0.1em\hbox{\small \it MB}}}
\newcommand{\Smb}{S_{\kern-0.1em\hbox{\small \it m-b}}}

\def\eff{\hbox{\small\it eff}\,}
\def\SeffU{S_{\kern -0.1em \eff}[u]}

\def\eqref#1{(\ref{#1})}

\title{Chiral-Soliton Predictions for Exotic Baryons}
\author{\strut\vskip-1cm
John Ellis\\
Theory Division, CERN, Geneva, Switzerland\\
E-mail: \email{john.ellis@cern.ch}}
\author{Marek Karliner\\
Cavendish Laboratory, Cambridge University, England \\
\kern3em and \\
School of Physics and Astronomy,\\
Raymond and Beverly Sackler Faculty of Exact Sciences,\\
Tel Aviv University, Tel Aviv, Israel\\
E-mail: \email{marek@proton.tau.ac.il
\vskip-0.6cm\strut
}}
\author{Micha{\l} Prasza{\l}owicz\\
Nuclear Theory Group, Brookhaven Nat. Lab., Upton, USA\\
\kern3em and \\
M. Smoluchowski Institute of Physics,
Jagellonian University, Krak{\'o}w, Poland\\
E-mail \email{michal@quark.phy.bnl.gov
\vskip-0.8cm}}

\preprint{CERN-TH/2003-288\\
Cavendish-HEP-04/04\\
BNL-NT-04/4\\
TPJU-1/2004}


\abstract{
We re-analyze the predictions of chiral-soliton models for the masses
and decay widths of baryons in the exotic antidecuplet of flavour
SU(3). The calculated ranges of the chiral-soliton moment of inertia and
the $\pi$-nucleon scattering $\SigmapN$ term are used together with the
observed baryon octet and decuplet mass splittings to estimate $1430~{\rm 
MeV} < m_{\Theta^+} < 1660~{\rm MeV}$ and $1790~{\rm MeV} < m_{\Xi^{--}} <
1970~{\rm MeV}$. These are consistent with the masses reported recently,
but more precise predictions rely on ambiguous identifications of
non-exotic baryon resonances. The overall decay rates of antidecuplet
states are sensitive to the singlet axial-current matrix element in the
nucleon. Taking this from polarized deep-inelastic scattering experiments,
we find a suppression of the total $\Theta^+$ and $\Xi^{--}$ decay widths
that may not be sufficient by itself to reproduce the narrow widths
required by experiments. We calculate $SU(3)$ breaking effects due to
representation mixing and find that they tend to suppress the $\Theta^+$
decay width, while enhancing that of the $\Xi^{--}$. We predict light
masses for some exotic $27$ baryons, including the $I = 1, J^P = {3
\over 2}^+$ $\Theta^+$ and $I = {3\over2}, J^P = {3 \over 2}^+$ $\Xi$
multiplets, and calculate their decay widths.
\vskip-1cm$\phantom{a}$
}

\begin{document}

\section{Introduction}

The constituent-quark model (CQM) has long reigned supreme as the
default approach to hadron structure, masses and
decays~\cite{CQM}. However, the CQM for light quarks has never
been derived from QCD, and a complementary point of view is
expressed in the chiral-soliton model
(\chiSM)~\cite{chiSM1,chiSM2}. Motivated by the observation that
the short-distance current masses of the $u, d$ and $s$ quarks are
all $\lappeq 100$~MeV, and the suggestion that chiral SU(3)
$\times$ SU(3) may be a good symmetry of hadrons, the \chiSM\
treats baryons as topologically non-trivial configurations of the
pseudoscalar meson fields. In some versions constituent quarks are
considered to have been integrated out of the effective
Lagrangian, whose solitons are interpreted as baryons. In the
original form of the \chiSM\ proposed by Skyrme~\cite{Skyrme} the
effective Lagrangian only contained terms up to quartic in the
derivatives of the meson fields, a restriction we do not apply
here. In other versions constituent quarks are explicitly
present~\cite{chiSM1,chiSM2}, but the effective baryon theory has
the same group-theoretical structure as purely mesonic models.

Despite considerable theoretical support and phenomenological exploration,
the \chiSM\ has remained a minority interest, probably because it lacks
the intuitive appeal and many of the phenomenological successes of the
CQM. However, the \chiSM\ has its own  successes, such as its
prediction of 5/9 for the $F/D$ ratio for axial-current nucleon matrix
elements - which is arguably more successful than the CQM prediction of
2/3, the Guadagnini relation~\cite{Guad} between flavour SU(3) symmetry
breaking in the lowest-lying baryon octet and decuplet, and the prediction
that the singlet axial-current nucleon matrix element should be
small~\cite{BEK}.

The CQM and \chiSM\ are to a large degree complementary. Each of them
reproduces certain aspects of hadronic physics and incorporates many
features of QCD which are missing in the other. This complementarity
between the CQM and \chiSM\
has some analogies to the relationship between the
shell model and the droplet model of the atomic nucleus. Neither provides a
complete description of the nucleus, but each one has its strengths.
Faced with a new phenomenon, one should therefore try to understand it
within each of the two approaches, and the best understanding may come from
combining features of both approaches.

For a long time, a potential embarrassment for the \chiSM\ has been its
prediction of exotic baryons. Beyond the lowest-lying $J = I = 1/2, 3/2$
baryons, the simple-minded SU(2) \chiSM\ predicted a tower of heavier $J
= I = 5/2, 7/2, ...$ states, which have never been seen. However, the
picture in the SU(3) version of the \chiSM\ is rather different: in this
framework, the lowest-lying exotic baryon is an antidecuplet ${\overline
{10}}$~\cite{Guad,antidec}--\nocite{DPP,Weigel}\cite{Prasz},
with other exotic representations such as the $27,
35$ and ${\overline {35}}$ being heavier. For most of two decades, the
existence of a light baryon antidecuplet has been a key unverified
prediction of the \chiSM.

This exotic `bug' may recently have turned into a feature,
following the discovery of the exotic $\Theta^+ (1540)$
baryon~\cite{Theta,others} with a relatively low mass and small
decay width as predicted in the \chiSM~\cite{DPP,Prasz,BieDo}.
However, alternative postdictive interpretations of this state
abound, including CQM descriptions~\cite{KL,JW,otherq},
kaon-baryon molecules~\cite{KB}, kaon-Skyrmion bound
states~\cite{Kleb}, etc. The \chiSM\ had also been used to predict
the masses of the other baryons in
the ${\overline {10}}$, including a $\Xi^{--}$ with mass between
2070~MeV~\cite{DPP} and $\sim 1850$~MeV \cite{Prasz,Kopel}, in
strong correlation with the assumed value of the pion-nucleon
sigma term. On the other hand the CQM approaches predicted
$m_{\Xi^{--}} \lappeq 1760$~MeV~\cite{KL,JW}. The NA49
collaboration has recently reported the observation of a candidate
$\Xi^{--}$ with a mass $\simeq 1860$~MeV~\cite{Xi}, within this
range of predictions. As we show below, a careful re-analysis of
the results of Ref.\cite{DPP} yields a range for the $\Xi^{--}$
mass that includes the the experimental value.\footnote{The
relation of NA49 result to previous data is discussed in
\cite{Fischer:2004qb}.}

The purpose of this paper is to
discuss critically the predictions for
the exotic $\Theta(1540)^+$ and $\Xi_{\overline{10}}$ baryons within the \chiSM.
As we discuss below, the masses and
decays of the exotic ${\overline {10}}$ baryons are uniquely sensitive to
the baryonic matrix elements of the SU(3)-singlet combinations of scalar
${\bar q} q$ densities and of axial ${\bar q} \gamma_\mu \gamma_5 q$
currents, respectively, and we discuss carefully the implications for
exotic baryons of the experimental uncertainties in these quantities.

In order to predict the masses of the $\Theta^+$ and $\Xi^{--}$,
we use estimates of the chiral-invariant contributions from
specific \chiSM\ calculations~\cite{Prasz,moments}, and estimates
of the chiral SU(3) $\times$ SU(3) symmetry breaking contributions
based the masses of octet and decuplet baryons and the $\SigmapN$ 
term in $\pi$-nucleon scattering~\cite{Sigma}. Using the range
$0.43~{\rm fm} < I_2 < 0.55~{\rm fm}$~\cite{Prasz,moments} for the
chiral-soliton moment of inertia that characterizes the difference
between the ${\overline {10}}$ and $10$ masses in the chiral
limit, and the range $64~{\rm MeV} < \SigmapN < 79~{\rm MeV}$ for 
the $\pi$-nucleon $\SigmapN$ term~\cite{Sigma}, we find the 
following ranges:
$1430~{\rm MeV} < m_{\Theta^+} < 1660~{\rm MeV}$ 
and $1790~{\rm MeV} < m_{\Xi^{--}} < 1970~{\rm MeV}$~\footnote{The 
considerably larger value of $I_2$ advocated 
in~\cite{moments} would yield ${\overline {10}}$ masses that were 
unacceptably light, and specific model calculations correlate the values 
of $I_2$ and $\SigmapN$.}. The more 
specific predictions made previously~\cite{DPP} relied on 
identifications of other resonances that are questionable, and/or 
a different value for the $\pi$-nucleon $\SigmapN$ term. We use 
values of \chiSM\ parameters inferred from the $\Theta^+$ and 
$\Xi^{--}$ masses to predict the masses of low-lying exotic
baryons in a $J^P = {3 \over 2}^+$ $27$ representation of flavour
SU(3)~\cite{Kopel,other27}, and calculate their decay widths.

In order to
predict the decay rates of the $\Theta^+$ and $\Xi^{--}$, one needs to
know a specific combination of the octet and singlet axial-current matrix
elements in the nucleon octet. In the absence of SU(3) symmetry breaking,
in the leading order of the $1/N_c$ expansion,
the \chiSM\ would predict that the singlet axial-current matrix element
vanishes~\cite{BEK}, in qualitative agreement with measurements of
polarized deep-inelastic lepton-nucleon scattering. However,
the deep-inelastic data indicate a small but non-zero singlet
axial-current matrix element, which is accommodated by $1/N_c$ and
${\cal O}(m_s/\Lambda_{QCD})$
corrections in the \chiSM~\cite{Christov}. Inserting the value of the
singlet
axial-current matrix element extracted from polarized deep-inelastic
lepton scattering experiments into the \chiSM\ formulae reduces somewhat
the decay widths of the $\Theta^+$ and $\Xi^{--}$, but perhaps
not sufficiently to
explain alone the very narrow widths of these states that are indicated by
experiment \cite{Theta,others,ThetaWidth}.
Representation mixing introduces $SU(3)$ breaking effects that suppress
the $\Theta^+$ decay width, while enhancing that of $\Xi^{--}$. They also
have an important effect on the $\pi$-nucleon
coupling that we calculate as well.

\section{Review of Relevant Aspects of the \chiSM}

We recall that the splittings between the centres of the lowest-lying
octet, decuplet and antidecuplet baryons are given in the \chiSM\ by
\begin{equation}
\Delta M_{{10} - 8} \; = \; {3 \over 2 I_1}, \; \;
\Delta M_{{\overline {10}} - 8} \; = \; {N_c \over 2 I_2}
\; = \; {3 \over 2 I_2}
\label{DeltaM}
\end{equation}
where $I_{1,2}$ are two soliton moments of inertia that depend on details
of the chiral Lagrangian.
Since $I_1$, $I_2\sim {\cal O}(N_c)$, this means that
$\Delta M_{{\overline {10}} - 8} \sim {\cal O}(N_c^0)$, whereas
$\Delta M_{{10} - 8}$ is  ${\cal O}(1/N_c)$. This has triggered some
arguments~\cite{Cohen} and counter-arguments~\cite{DP9}, regarding the
applicability of collective coordinate
quantization to the $\overline {10}$. We note here that the application
of the collective quantization relies on the rotor excitation being small
in comparison with the classical mass. Since the latter is ${\cal O}(N_c)$,
this requirement holds for ${\overline {10}}$ as well, even though
the suppression is just ${\cal O}(1/N_c)$ vs. ${\cal O}(1/N_c^2)$ for
the {10} and the {8}. Experimentally, $\Delta M_{{10} - 8}=231$ MeV
whereas $\Delta M_{{\overline {10}} - 8} \sim 600$ MeV, in good agreement
with formal $N_c$ counting \footnote{Provided we interpret both the
$\Theta^+$ and the $\Xi^{--}$ as members of the $\overline{10}$
multiplet.}.

The centre of the lightest octet of baryons is
the average of the $\Lambda$ and $\Sigma$ masses, namely 1151.5~MeV, and
the centre of the $10$ of baryons is that of the $\Sigma_{10}$, namely
1382.1~MeV~\cite{PDG}. The centre of the ${\overline {10}}$ would likewise
be identified with the $\Sigma_{\overline {10}}$, which may mix in general
with the $\Sigma_8$ expected in the same band of soliton excitations, and
even with other adjacent $\Sigma_8$ states. Analogous mixing is expected
for the $N_{\overline {10}}$. In the pioneering analysis of~\cite{DPP},
the known $N(1710)$ was identified with the $N_{\overline {10}}$.
However, such identifications are ambiguous, since the baryon
spectrum is expected to contain both radial and rotational excitations
that mix in general~\cite{Weigel,Cohen,DP9}.
These identifications were abandoned in~\cite{DP}.
In this paper, we do not impose the identification of the $N(1710)$ or any
other known nucleon resonance such as the $N(1440)$ with any combination
of the solitonic $N_{{\overline {10}}, 8}$ states.

The leading-order chiral-symmetry breaking corrections to the lightest
octet baryon masses are~\cite{DPP}:
\begin{eqnarray}
N \; &:& \; + {3 \over 10} \alpha + \beta - {1 \over 20} \gamma, \\
\Lambda \; &:& \; + {1 \over 10} \alpha + {3 \over 20} \gamma, \\
\Sigma \; &:& \; - {1 \over 10} \alpha - {3 \over 20} \gamma, \\
\Xi \; &:& \; - {1 \over 5} \alpha - \beta + {1 \over 5} \gamma,
\label{8masses}
\end{eqnarray}
where the parameters $\alpha, \beta, \gamma$ cannot now be
determined from first principles. In particular, $\beta$ and
$\gamma$ are related to ratios of soliton moments of
inertia~\cite{Prasz,moments}:
\begin{equation}
\beta \; = \; -m_s {K_2 \over I_2}, \;
\gamma \; = \; 2 m_s \left( {K_1 \over I_1} - {K_2 \over I_2}
\right),
\label{DPPbeta}
\end{equation}
These origins impose on them some positivity conditions,
namely: \begin{equation}
\beta \; < \; 0, \; {1 \over 2} \gamma - \beta > 0.
\label{positivity}
\end{equation}
We also note that $\beta$ and $\gamma$ are formally of higher order in
$1/N_c$ than $\alpha$, and hence should be somewhat smaller than $\alpha$.
Whatever the values of $\alpha, \beta$ and $\gamma$, the octet baryons
should obey the Gell-Mann-Okubo mass formula
\begin{equation}
2 (m_N + m_\Xi) \; = \; 3 m_\Lambda + m_\Sigma,
\label{GMO}
\end{equation}
which is quite well satisfied experimentally. In the case of the decuplet
baryons, one has the leading-order mass corrections
\begin{eqnarray}
\Delta \; &:& \; + {1 \over 8} \alpha + \beta - {5 \over 16} \gamma, \\
\Sigma^* \; &:& \;\;\;\; 0, \\
\Xi^* \; &:& \; - {1 \over 8} \alpha - \beta + {5 \over 16} \gamma, \\
\Omega \; &:& \; - {1 \over 4} \alpha - 2 \beta + {5 \over 8} \gamma,
\label{10masses}
\end{eqnarray}
which provide the standard equal-spacing mass formula for the $10$
multiplet:
\begin{equation}
m_{\Sigma^*} - m_\Delta \; = \; m_{\Xi^*} - m_{\Sigma^*} \; = \; m_\Omega
- m_{\Xi^*},
\label{equalmass}
\end{equation}
which is also quite well satisfied~\footnote{We comment later on the
potential significance of corrections of higher order in SU(3) symmetry
breaking~\cite{quadratic,Prasz}.}. As a bonus, one
obtains the Guadagnini relation~\cite{Guad} between the $8$ and $10$ mass
splittings:
\begin{equation}
8 (m_{\Xi^*} + m_N) + 3 m_\Sigma \; = \; 11 m_\Lambda + 8 m_{\Sigma^*},
\label{Guad}
\end{equation}
which is satisfied almost as accurately as (\ref{GMO}) and
(\ref{equalmass}). Finally, in the case of the ${\overline {10}}$ baryons,
one has the mass corrections
\begin{eqnarray}
\Theta^+ \; &:& \; + {1 \over 4} \alpha + 2 \beta - {1 \over 8} \gamma, \\
N_{\overline {10}} \; &:& \; + {1 \over 8} \alpha + \beta - {1 \over 16}
\gamma, \\
\Sigma_{\overline {10}} \; &:& \;\;\;\; 0, \\
\Xi_{I = 3/2} \; &:& \; - {1 \over 8} \alpha - \beta + {1 \over 16}
\gamma,
\label{10barmasses}
\end{eqnarray}
which also leads to equal spacings, but with magnitudes different from
those in the decuplet of
baryons~\footnote{We note in passing that the CQM
also predicts equal spacing for the ${\overline {10}}$ baryons,
but different from that for the ordinary decuplet:
$\Delta M_{\overline{10}}\,\sim(m_s-m_u)/3$,
before
the possible mixing of the $N_{8, {\overline {10}}}$ and $\Sigma_{8,
{\overline {10}}}$.}.

Reflecting the existence of the Guadagnini mass relation (\ref{Guad}), we
recall that the mass corrections (2) to (5) and (9) to (12) depend on just
two combinations of the parameters $\alpha, \beta$ and $\gamma$, which
may be determined as follows in a least-squares fit:
\begin{eqnarray}
\alpha + {3 \over 2} \gamma \; & = & \; {-}377~{\rm MeV}, \\
{1 \over 8} \alpha + \beta - {5 \over 16} \gamma \; & = & \;{-}146~{\rm
MeV}.
\label{2for3}
\end{eqnarray}
A third relation is necessary if one is to determine $\alpha, \beta$ and
$\gamma$ and calculate the mass corrections (14) to (17). This can be
provided by the chiral-symmetry breaking expression for the $\sigma$ term in
$\pi$-nucleon scattering:
\begin{equation}
\alpha + \beta \; = \; - {2 \over 3} {m_s \over m_u + m_d} \Sigma\,
\label{Sigmaterm}
\end{equation}
where baryon and meson data yield the estimate $m_s / (m_u + m_d) =
12.9$~\cite{Leutwyler}. 
As is well-known, the value of $\SigmapN$ is related to the nucleon matrix
element of the SU(3)-singlet combination $\langle N | ( {\bar u} u +
{\bar d} d + {\bar s} s) | N \rangle$.

\section{Predictions for the Masses of Antidecuplet Baryons}

We have seen that, to predict the masses of the $\Theta^+$ and
$\Xi^{--}$, one must obtain values for the soliton moment of
inertia $I_2$ and the chiral-symmetry breaking parameters $\alpha,
\beta$ and $\gamma$. Different soliton models yield values for
$I_2$ in the range~\cite{Prasz,moments}:
\begin{equation}
0.43~{\rm fm} \; < \; I_2 \; < \; 0.55~{\rm fm},
\label{I2}
\end{equation}
which yields the range
\begin{equation}
538~{\rm MeV} \; < \; \Delta M_{{\overline {10}} - 8} \; < \; 638~{\rm
MeV}.
\label{DeltaMnumbers}
\end{equation}
In the version of the Skyrme model discussed in~\cite{chiSM1} the upper 
limit on $I_2$ is even higher, being 
of the order of 1 fm. We note, however, that such a value of $I_2$ would 
bring the $\overline{10}$ masses unacceptably low.

 To determine the chiral-symmetry breaking corrections, we
use the central values of two recent determinations of the
$\pi$-nucleon $\Sigma$ term: $\SigmapN = 64 \pm 8$ $(79 \pm 7)$ 
MeV~\cite{Sigma}. We now have three equations for the three
unknowns $\alpha, \beta$ and $\gamma$, for which we find the
values needed to predict ${\overline {10}}$ baryon masses in the
\chiSM:
\def\MeV{{\rm MeV}}
\def\mystrut{\vrule height 2.5ex depth 0ex width 0pt}
\bigskip
\begin{equation}
\begin{array}{c|rc|rc}
\SigmapN &  64 \pm 8\,\,\,\,\,\,   & \MeV & 79 \pm 7\,\,\,  & \MeV\\
\hline
\mystrut
\alpha & {-}489 \pm103            & \MeV& {-}683\pm90           & \MeV \\
\beta  & {-}61  \pm\phantom{1}34  & \MeV& \phantom{{-}}\phantom{68}3\pm30 & \MeV \\
\gamma &  74    \pm\phantom{1}69  & \MeV& \phantom{{-}}203\pm60 & \MeV
\label{CSBnumbers}
\end{array}
\end{equation}
\mystrut\hfill\break
Using the ranges (\ref{DeltaMnumbers}, \ref{CSBnumbers}), we find the
following ranges for the masses of the exotic baryons in the ${\overline
{10}}$ multiplet:
\begin{eqnarray}
{\rm for}\ \SigmapN=64\ {\rm MeV}: \quad
m_{\Theta^+}=1505^{{+}84}_{{-}66}\,\ {\rm MeV}, 
\quad
m_{\Xi^{--}}=1885^{{+}84}_{{-}66}\,\ {\rm MeV}
\\
\mbox{} \nonumber
\\
{\rm for}\ \SigmapN=79\ {\rm MeV}: \quad
m_{\Theta^+}=1569^{{+}84}_{{-}66}\,\ {\rm MeV}, 
\quad
m_{\Xi^{--}}=1853^{{+}84}_{{-}66}\,\ {\rm MeV} 
\label{newpreds}
\end{eqnarray}
where the upper and lower errors reflect the variation of $I_2$,
eq.~\eqref{I2}. An additional error comes from the ${\sim}7$ MeV
uncertainty in the central values of $\SigmapN$: 
$\delta m_{\Theta^+}/\delta \SigmapN\approx 4$,
$\delta m_{\Xi^{--}}/\delta \SigmapN\approx 2$. Overall, we find the ranges
\begin{equation}
1432~{\rm MeV} < m_{\Theta^+} < 1657~{\rm MeV}, \quad
1786~{\rm MeV} < m_{\Xi^{--}} < 1970~{\rm MeV},
\label{summarypreds}
\end{equation}
upon combining these errors in quadrature.

The ranges \eqref{summarypreds} certainly include the observed masses 
$m_{\Theta^+}{=}1539{\pm}2$~{\rm MeV} and $m_{\Xi^{--}} = 1862 \pm 2$~{\rm 
MeV}, but more precise predictions cannot be made without 
introducing more assumptions. In our view, the success of the prediction 
of~\cite{DPP} for the $\Theta^+$ mass was somewhat fortuitous. 
Ref.~\cite{DPP} identified the $N_{\overline {10}}$ with the $N(1710)$, 
and assumed an older value for the $\pi$-nucleon $\Sigma$ term: $\SigmapN 
= 45$~MeV. It is this latter value, in particular, that was responsible 
for the unsuccessful prediction in~\cite{DPP} of a very heavy mass $\sim
2070$~MeV for the $\Xi^{--}$ state. 
This conclusion is reinforced by the analysis of 
Ref.~\cite{Schweitzer:2003fg} where $\SigmapN=74\pm12$ MeV is obtained 
from the observed spectrum of usual and exotic baryons.

The ratio of $m_s/(m_u+m_d)=12.9$ that we have assumed in the
above analysis corresponds to the strange quark mass
$m_s=140$~MeV for $m_u+m_d=11$~MeV. It is, however, quite
possible that quark masses in the effective models take values
different than in the underlying QCD theory. In~\cite{moments} the best fit
value of the strange quark mass was approximately 185 -- 195~MeV, rather 
than 140~MeV. Since $m_s$ and $\SigmapN$ enter as a product into
(\ref{Sigmaterm}), one can compensate the large value of the
latter by increasing $m_s$. This would introduce another 25\%
uncertainty into the estimates (\ref{summarypreds}). In what
follows, we do not use the value of the $\SigmapN$ term
any more, but fit the model parameters to the measured baryon
masses.

Within the \chiSM\ framework, the observed masses of the $\Theta^+$ and 
$\Xi^{--}$ can be used, together with the masses of ordinary 
octet and decuplet, to estimate the key model parameters,
whose interpretation we discuss below:
\begin{equation}
I_1=1.27~{\rm fm},\;
I_2  =0.49~{\rm fm}, \;
\alpha = {-}605~{\rm MeV}, \; \beta = {-}23~{\rm MeV},
 \;
\gamma =  152~{\rm MeV},
\label{exptnumbers}
\end{equation}
corresponding to $\SigmapN = 73$ MeV. We see again that the reported mass of 
the $\Xi_{\overline{10}}$ is no problem for the \chiSM. 
Having fixed all the parameters 
of the model we predict the remaining $\overline{10}$ masses: $M_{N^{\ast}}= 
1646$~MeV and $M_{\Sigma}=1754$~MeV.

We now check the consistency of the leading-order
expansion in SU(3) symmetry breaking, by incorporating the representation
mixing due to the SU(3)-breaking Hamiltonian:
\begin{equation}
\hat{H}^{\prime}=\alpha D_{88}^{(8)}+\beta Y+\frac{\gamma}{\sqrt{3}}%
D_{8i}^{(8)}\hat{S}_{i},
\label{Hprime}
\end{equation}
Most relevant for this paper are the following mixings induced by
(\ref{Hprime}):
\begin{align}
\left|  B_{8}\right\rangle  &  =\left|  8_{1/2},B\right\rangle +c_{\overline
{10}}^{B}\left|  \overline{10}_{1/2},B\right\rangle +c_{27}^{B}\left|
27_{1/2},B\right\rangle ,\nonumber\\
\left|  B_{10}\right\rangle  &  =\left|  10_{3/2},B\right\rangle +a_{27}%
^{B}\left|  27_{3/2},B\right\rangle +a_{35}^{B}\left|  35_{3/2},B\right\rangle
,\nonumber\label{states}\\
\left|  B_{\overline{10}}\right\rangle  &  =\left|  \overline{10}%
_{1/2},B\right\rangle +d_{8}^{B}\left|  8_{1/2},B\right\rangle +d_{27}%
^{B}\left|  27_{1/2},B\right\rangle +d_{\overline{35}}^{B}\left|
\overline{35}_{1/2},B\right\rangle
\end{align}
where
\begin{align}
c_{\overline{10}}^{B}  &  =c_{\overline{10}}\left[\kern-0.5em
\begin{array}
[c]{c}%
\sqrt{5}\\
0\\
\sqrt{5}\\
0
\end{array}
\kern-0.2em\right]\kern-0.2em,\;c_{27}^{B}=c_{27}\left[\kern-0.5em
\begin{array}
[c]{c}%
\sqrt{6}\\
3\\
2\\
\sqrt{6}%
\end{array}
\kern-0.2em\right]\kern-0.2em  ,\;a_{27}^{B}=a_{27}\left[\kern-0.5em
\begin{array}
[c]{c}%
\sqrt{15/2}\\
2\\
\sqrt{3/2}\\
0
\end{array}
\kern-0.2em\right] \kern-0.2em ,\;a_{35}^{B}=a_{35}\left[\kern-0.5em
\begin{array}
[c]{c}%
5/\sqrt{14}\\
2\sqrt{5/7}\\
3\sqrt{5/14}\\
2\sqrt{5/7}%
\end{array}
\kern-0.2em\right] \nonumber\\
d_{8}^{B}  &  =d_{8}\left[
\begin{array}
[c]{c}%
0\\
\sqrt{5}\\
\sqrt{5}\\
0
\end{array}
\right]  ,\qquad d_{27}^{B}=d_{27}\left[
\begin{array}
[c]{c}%
0\\
\sqrt{3/10}\\
2/\sqrt{5}\\
\sqrt{3/2}%
\end{array}
\right]  ,\qquad d_{\overline{35}}^{B}=d_{\overline{35}}\left[
\begin{array}
[c]{c}%
1/\sqrt{7}\\
3/(2\sqrt{14)}\\
1/\sqrt{7}\\
\sqrt{5/56}%
\end{array}
\right]  \label{mix}%
\end{align}
in the basis $[N,\Lambda,\Sigma,\Xi]$, $[\Delta,\Sigma^{\ast},\Xi^{\ast
},\Omega]$ and $\left[  \Theta^{+},N_{\overline{10}},\Sigma_{\overline{10}%
},\Xi_{\overline{10}}\right]  $ respectively, and
\begin{align}
c_{\overline{10}}=-\frac{I_{2}}{15}\left(  \alpha+\frac{1}{2}\gamma\right)  ,
&  ~c_{27}=-\frac{I_{2}}{25}\left(  \alpha-\frac{1}{6}\gamma\right)
,\nonumber\\
~~  & \nonumber\\
a_{27}=-\frac{I_{2}}{8}\left(  \alpha+\frac{5}{6}\gamma\right)  ,  &
~a_{35}=-\frac{I_{2}}{24}\left(  \alpha-\frac{1}{2}\gamma\right)  ,\nonumber\\
~~  & \nonumber\\
d_{8}=\frac{I_{2}}{15}\left(  \alpha+\frac{1}{2}\gamma\right)  ,~~  &
d_{27}=-\frac{I_{2}}{8}\left(  \alpha-\frac{7}{6}\gamma\right)
,~~~d_{\overline{35}}=-\frac{I_{2}}{4}\left(  \alpha+\frac{1}{6}\gamma\right)
. \label{mix1}%
\end{align}
For our set of parameters (\ref{exptnumbers})
the mixing coefficients range
from $0.06$ to $0.36$:
\begin{align}
c_{\overline{10}}  &  =-d_{8}=0.088,\;c_{27}=0.063,\;\nonumber\\
a_{27}  &  =0.150,\;a_{35}=0.071,\;\nonumber\\
d_{27}  &  =0.245,\;d_{\overline{35}}=0.362 \label{mixnum}%
\end{align}
which by (\ref{mix}) results in admixtures which are typically of the
order of $10$ to $20$\%.

These first-order admixtures lead to the following second-order
corrections to the masses of $8$, $10$ and
${\overline {10}}$ baryons~\cite{Prasz}:
\begin{align}
E_{8}^{(2)}(Y,T)  & =-I_{2}\left[  \frac{1}{60}\left(  Y+\left(
T(T+1)-\frac{1}{4}Y^{2}\right)  +\frac{1}{2}Y^{2}\right)  \left(
\alpha+\frac{1}{2}\gamma\right)  ^{2}\right. \nonumber\\
& \quad\left.  +\frac{1}{250}\left(  9-\frac{5}{2}\left(  T(T+1)-\frac{1}%
{4}Y^{2}\right)  -\frac{7}{4}Y^{2}\right)  \left(  \alpha-\frac{1}{6}%
\gamma\right)  ^{2}\right] \nonumber\\
& =-I_{2}\left[  \frac{1}{60}\left(  Y+T\left(  T+1\right)  +\frac{1}{4}%
Y^{2}\right)  \left(  \alpha+\frac{1}{2}\gamma\right)  ^{2}\right.
\nonumber\\
& =-I_{2}\left[  \frac{1}{60}\left(  Y+T\left(  T+1\right)  +\frac{1}{4}%
Y^{2}\right)  \left(  \alpha+\frac{1}{2}\gamma\right)  ^{2}\right.
\nonumber\\
& \left.  +\frac{1}{250}\left(  9-\frac{5}{2}T\left(  T+1\right)
-\frac{9}%
{8}Y^{2}\right)  \left(  \alpha-\frac{1}{6}\gamma\right)  ^{2}\right],
\end{align}
\begin{align}
E_{10}^{(2)}(Y)  & =-I_{2}\left[  \frac{1}{16}\left(  1+\frac{3}{4}%
Y+\frac{1}{8}Y^{2}\right)  \left(  \alpha+\frac{5}{6}\gamma\right)
^{2}\right. \nonumber\\
& \quad\left.  +\frac{5}{336}\left(
1-\frac{1}{4}Y-\frac{1}{8}Y^{2}\right)
\left(  \alpha-\frac{1}{2}\gamma\right)  ^{2}\right]  ,
\end{align}
\begin{align}
E_{\overline{10}}^{(2)}(Y)  & =I_{2}\left[  \frac{1}{30}\left(
1+\frac{1}%
{2}Y-\frac{1}{2}Y^{2}\right)  \left(  \alpha+\frac{1}{2}\gamma\right)
^{2}\right. \nonumber\\
& \qquad-\frac{1}{640}\left(  8-6Y+Y^{2}\right)  \left(  \alpha-\frac{7}%
{6}\gamma\right)  ^{2}\nonumber\\
& \qquad\left.  -\frac{3}{896}\left(  8+2Y-Y^{2}\right)  \left(
\alpha+\frac{1}{6}\gamma\right)  ^{2}\right]  .
\end{align}
We find, in particular, the following dominant second-order corrections to
the $\Theta^+$ and $\Xi_{\overline{10}}$ masses due to mixing with other
exotic rotational
excitations:
\begin{eqnarray}
\delta_2 m_{\Theta^+} & = & - {3 \over 112} I_2
\left( \alpha+\frac{1}{6}\gamma \right)^2, \;
\nonumber \\
\delta_2 m_{\Xi^{--}} & = &
-I_2\,\left( \frac{3}{128}
       \left( \alpha  -
           \frac{7}{6}\gamma \right)^2
        + \frac{15}{896}
       \left( \alpha  +
           \frac{1}{6}\gamma  \right)^2
    \right)
\label{Prasz}
\end{eqnarray}
where the effect comes from mixing with similar states in a
$27$ multiplet (related to $(\alpha-7\gamma/6)$) and a
${\overline {35}}$ multiplet (related to $(\alpha+\gamma/6)$).
Using the values of $\alpha$ and $\gamma$ extracted above
(\ref{exptnumbers}),
these corrections amount numerically to
${-}22.5$, ${-}50$~MeV, respectively.

It is likely that there are similar mass corrections due to mixing
with other states such as radial excitations~\cite{Cohen}. Some of
these mixings have been considered in~\cite{Weigel} in a specific model, 
but there could be additional effects of this type which
have not yet been fully investigated in the
literature~\footnote{Such mixing is likely to be more important
for non-exotic baryons, which is one reason why we do not advocate
estimating $I_2$ from fits to baryon masses including quadratic
corrections.}. If included in the above fit to the exotic baryon
masses, the corrections (\ref{Prasz}) would correspond to shifting
$I_2 \to 0.51$~fm and $\SigmapN \to 72$~MeV. These small
changes indicate that the procedure~\cite{DPP} of calculating mass
corrections to first order in $1/N_c$ and SU(3) symmetry breaking
may be reasonably stable.

It is reassuring to note that the extracted values of $\alpha$ and $\beta$ 
correspond to a value of $\SigmapN$ between the two recent experimental 
determinations~\cite{Sigma}. 
We also note that the extracted values of $\beta$ and $\gamma$ 
(\ref{exptnumbers}) respect the 
positivity constraints (\ref{positivity}) required in the \chiSM, and 
that $| \alpha | \gg | \beta |, | \gamma |$, as expected on the basis of 
the $1/N_c$ expansion. Inserting the value $\beta = 
{-}23$~MeV (\ref{exptnumbers}), extracted from the masses of the known 
antidecuplet states, the \chiSM\ expression for $\beta$ (\ref{DPPbeta}) 
and the estimate $m_s \sim 100 - 200$~MeV suggest the following value for 
the ratio of two moments of inertia:
\begin{equation}
{K_2 \over I_2} \; = \; 0.23 - 0.11,
\label{K2value}
\end{equation}
which is quite small. However, a realistic error on $\beta$ might be
35~MeV, in which case somewhat larger values of $K_2$ would also be
possible. We note also that (\ref{DeltaM}) and the observed octet and
decuplet masses yield $I_1 = 1.29$~fm. Considering now the
\chiSM\ expression for $\gamma$ (\ref{DPPbeta}),
we see that the small ratio $K_2/I_2$ (\ref{K2value}) is quite consistent
with the positive value of $\gamma$ found in (\ref{exptnumbers}), and
yields the following value for the ratio of two other moments of inertia in the
\chiSM:
\begin{equation}
{K_1 \over I_1} \; = 0.98 - 0.49
\label{K1value}
\end{equation}
for $m_s = 100 - 200$~MeV. The extracted values of the four
moments of inertia $I_{1,2}$ and $K_{1,2}$ provide interesting constraints on
the \chiSM\ that lie beyond the scope of this paper, though we note that
they are not typical of model calculations.

\section{Predictions for the Decay Widths of Exotic Baryons}

\subsection{General Remarks on Decay Widths in the \chiSM\ }

Whilst the mass spectra discussed in the previous Section are given as
systematic expansions in both $N_{c}$ and $m_{s}$ in a theoretically
controllable way, reliable predictions for the decay widths cannot be
organized in a similar manner. As explained below, they depend on
modelling and `educated' guesses, and hence are subject to additional
uncertainties.

The width for any decay $B\rightarrow B^{\prime}+\varphi$ may be expressed
in terms of the matrix element $\mathcal{M}$ and a two-body phase-space
factor:
\begin{equation}
\Gamma_{B\rightarrow B^{\prime}\varphi}=\frac{\overline{\mathcal{M}^{2}}}{8\pi
M\,M^{\prime}}p_{\varphi} \label{Gamdef1}%
\end{equation}
\hfill\break where $\varphi$ is a pseudoscalar meson with momentum
$p_{\varphi}$ in the $B$ rest frame:%
\[
p_{\varphi}=\frac{\sqrt{(M^{2}-(M^{\prime}-m_{\varphi})^{2})(M^{2}-(M^{\prime
}-m_{\varphi})^{2})}}{2M}%
\]
and the bar over ${\mathcal{M}}^2$ in (\ref{Gamdef1}) denotes an average
over the initial and a sum over the final spins, and -- when explicitly
indicated -- summing and averaging over isospin.

The first uncertainty comes from the fact that the baryon masses $M$ and
$M^{\prime}$ appear in the denominator of (\ref{Gamdef1}) yielding,
formally, {\it infinite series} in $N_{c}$ and $m_{s}$. The same holds
for the momentum of the outgoing meson $\varphi$. It is a common practice
to treat the phase factor exactly, rather than expand it up to a given
order in $N_{c} $ and $m_{s}$, despite the fact that in the matrix element
$\mathcal{M}$ only a few first terms in $1/N_{c}$ and $m_{s}/\Lambda_{QCD}$ are
calculated.

Secondly, $\mathcal{M}$ stands in (\ref{Gamdef1}) for the {\it
relativistic} matrix element which, in the case of nucleon decay, could be
calculated from the Lagrangian density considered already by
Adkins, Nappi and Witten in Ref.~\cite{Skyrme}:
\begin{equation}
\mathcal{L}_{int}=ig_{\pi NN}\,\pi^{a}\left(  \overline{\psi}\gamma_{5}%
\tau_{a}\psi\right)  . \label{Lintdens}%
\end{equation}
Unfortunately, we have at our disposal a \emph{non-relativistic} model, in
which baryons are considered as infinitely heavy, rather than a
relativistic field theory like (\ref{Lintdens}). It was already observed
in \cite{Skyrme} that the non-relativistic reduction of
(\ref{Lintdens}) leads to the interaction Lagrangian (as extended to
SU(3)):
\begin{equation}
L_{int}=g\,\partial^{i}\varphi^{\alpha}\,A_{\alpha i} \label{Lint},
\end{equation}
where $A_{i\alpha}$ is a spatial component of an axial current of flavour
$\alpha$. Here, $g$ is a coupling constant related to $g_{\pi NN}$ which
depends, in principle, on the initial and final baryon
states~\cite{DPP}. Furthermore, it is clear that the
Lagrangian density for spin-$3/2$ baryons decaying into baryons of spin
$1/2$ cannot be cast in the form (\ref{Lintdens}), because it must involve
a Rarita-Schwinger spinor which carries an extra vector index and has
different canonical dimension. Luckily, even in this case, one can still
use (\ref{Lint}), but the coupling constant $g$ should be appropriately
rescaled \cite{DPP}.

It is appropriate at this point to keep in mind the well-known and
difficult problem of 
Yukawa couplings in \chiSM (see eg. \cite{Yukawa} for an in-depth
discussion).
The fundamental source of this problem is
that baryons are constructed from meson fields and so in leading 
order in $1/N_c$ terms linear in
mesons vanish when one expands around the soliton configuration.

 There have been several interesting attempts to resolve this
problem, but at present there is no consensus about their effectiveness
Since we are focusing here on exotics, a detailed discussion
of this problem would take us much beyond the scope of the present paper.

The baryon decay operator following from (\ref{Lint}) can be written as
\begin{equation}
\hat{O}_{\varphi}^{(8)}=3\left[  G_{0}D_{\varphi i}^{(8)}-G_{1}d_{ibc}%
\,D_{\varphi b}^{(8)}\hat{S}_{c}-G_{2}\frac{1}{\sqrt{3}}D_{\varphi8}%
^{(8)}\hat{S}_{i}\right]  p_{\varphi\,\iota}, \label{Ophi}%
\end{equation}
which transforms as an octet of SU(3). The decay matrix elements
${\mathcal{M}}=\left\langle B^{\prime}\right|  \hat{O}_{\varphi}^{(8)}\left|
B\right\rangle $ may then be written in terms of the couplings
$G_{0,1,2}$, which are in turn related to
axial-current matrix elements $a_{0,1,2}$:
\begin{equation}
A_{\alpha i}=a_{0}D_{\alpha i}-a_{1}d_{ibc}\,D_{\alpha b}^{(8)}\hat{S}_{c}%
-\frac{a_{2}}{\sqrt{3}}D_{\alpha8}^{(8)}\hat{S}_{i} \label{curr}%
\end{equation}
by generalized Goldberger-Treiman relations:
\begin{equation}
G_{i}=ga_{i}.%
\label{gai}
\end{equation}
As discussed in~\cite{Dorey}, there are $m_s$ and
$1/N_c$ corrections to such relations associated with form factors in
axial-current matrix elements. Their calculation would involve a treatment
of deformations of soliton configurations, of which the principles and one
example are given in~\cite{Dorey}, but which have never been calculated in
a model-independent way. Here we use (\ref{gai}) and comment later on the
possible impact of deviations from it.

In what follows, we first make the approximation, following (\ref{Lint}),
that $g$ is a $\emph{universal}$ constant and calculate the decay widths
using (\ref{Gamdef1}), which is a good initial approximation, since
decuplet and antidecuplet decays are governed by two different linear
combinations of the coupling constants $G_{0,1,2}$, and, even using the
constraints from the semileptonic hyperon decays, one has enough freedom
to accommodate simultaneously large decuplet widths and small
$\overline{10}$ widths. However, when we also include the leading $m_{s}$
corrections, or try to estimate the $g_{\pi NN}$ coupling, or the
suppressed decay widths of the ${\overline {10}}$ baryons and the widths
of more exotic states such as those in the $27$ multiplets, then we need
to know $G_{0,1,2}$ {\em separately}. Then it becomes important whether
the corrections due to the mass dependence of $g$ are included or not.
Here, we include them following~\cite{DPP}, and discuss the potential
uncertainties in our predictions that they reflect.

At leading order in the $1/N_{c}$ expansion, corresponding to
ultra-non-relativistic baryons, all the couplings of the $8,10$ and
${\overline{10}}$ baryons to pseudoscalar mesons are proportional to the
dimensionless constant $G_{0}$ introduced above, with $F/D=5/9$
\cite{bijnens}. In this
approximation, the chiral soliton has no coupling to the singlet
pseudoscalar-meson field, and the singlet axial-current matrix elements
vanish. This provides a qualitative explanation~\cite{BEK} for the smallness
of the singlet axial-current matrix element of the nucleon inferred from
polarized deep-inelastic lepton scattering experiments.

The constants $G_{1,2}$ are non-leading as far as $N_{c}$ counting is
concerned. However, in antidecuplet decays the $G_{1}$ contribution gets
an additional $N_{c}$ enhancement from the SU(3)-flavour Clebsch-Gordan
coefficients calculated in large $N_{c}$ limit~\cite{Prasz11}.

One source of $m_{s}$ corrections is representation mixing. As already
discussed in connection with baryon masses, in the presence of SU(3)
breaking the physical states are no longer pure octet, decuplet or
antidecuplet states, but contain admixtures (\ref{states}) of the order of
$m_{s}$. Since their magnitudes are completely determined by the mass
splittings, their influence on the decay widths can be estimated reliably.
In the following, we use them below as estimates of the possible errors in
the decay widths associated with SU(3) symmetry breaking.

In addition to these calculable effects, the operator
$\hat{O}_{\varphi}^{(8)}$ gets additional $m_{s}$ corrections whose
algebraic structure is known from the analysis of semileptonic hyperon
decays~\cite{KPG}. These introduce three additional couplings, which we
ignore in the present phenomenological analysis, as their determinations
would require a lengthy analysis together with hyperon decays, which lies
beyond the scope of this paper.

\subsection{SU(3) Symmetry Limit} \label{symm}

Neglecting mixing between baryon multiplets, one has~\cite{Christov}
\begin{align}
g_{\pi NN}\; &  =\;\frac{7}{10}
\left[ (G_{0}+{\frac{1}{2}}G_{1}+{\frac{1}{14}}%
G_{2} \right],\label{gpinn}\\
G_{10} \; &=\;G_{0}+{\frac{1}{2}}G_{1},\label{gpiDN}\\
G_{\overline{10}}\; & =G_{0}-G_{1}%
-{\frac{1}{2}}G_{2}.\label{gKTN}%
\end{align}
which are related to $g_{\pi\Delta N}$ and $g_{K\Theta N}$,
$g_{\pi\Xi N}$ respectively. Furthermore
\begin{equation}
\frac{F}{D} ={\frac{5}{9}}  \; \left(
\frac{G_{0}+\frac{1}{2}G_{1}+\frac{1}{2}G_{2}}
     {G_{0}+\frac{1}{2}G_{1}-\frac{1}{6}G_{2}} \right).\label{FoverD}%
\end{equation}
The $G_{2}$ coupling is related \`{a} la Goldberger-Treiman to the singlet
axial-current matrix element in the nucleon:

\begin{equation}
G_{2}\;=\frac{2m_{N}}{3F_{\pi}}g_{A}^{0}, \label{GT0}%
\end{equation}
for $F_{\pi}=93$~MeV and may be non-zero. However, the consistency of the
$1/N_{c}$ expansion
would require $G_{2}$ to be relatively small, along with $G_{1}$.

To proceed further, we need input from $\Delta$ decay.
Using (\ref{gpiDN}), the measured decuplet decay width $\Gamma_{\Delta
}=115\div125$~MeV~\cite{PDG} and the theoretical prediction for the
$\Delta$ width
\[
\Gamma_{\Delta}=\frac{3G_{10}^{2}}{8\pi M\,_{\Delta}M_{N}}\frac{1}{5}\,p_{\pi
}^{3}%
\]
would yield the combination
\begin{equation}
G_{10}=G_{0}+{\frac{1}{2}}G_{1}\;=\;22.4.
\label{G0G1}%
\end{equation}
Unfortunately, we have no independent experimental information on any
other
combination of $G_{0}$ and $G_{1}$. Luckily, for non-exotic matrix
elements, we
only need $G_{10}$ and $G_{2}$. Therefore one finds
\begin{equation}
g_{\pi NN}\;=\frac{7}{10}G_{10}+\frac{G_{2}}{20}=15.6 +\frac{G_{2}}{20},
\label{G2from_gpinn}
\end{equation}
where the $G_2$-dependent correction is presumably small, in view of
(\ref{GT0}). The value (\ref{G2from_gpinn})
does not compare well with the experimental range $g_{\pi NN}=13.3\pm
0.1$ given in~\cite{Ericson} or the slightly different range $g_{\pi NN}
= 13.13 \pm 0.07$ recently advocated in~\cite{Bugg:2003bj},
and is not useful for extracting a numerical value of the
undetermined parameter $G_{2}$ wanted for calculating the decay
widths of the $\Theta^{+}$ and $\Xi^{--}$. Likewise, the baryon $F/D$ ratio
is not known sufficiently well to extract a useful value of
$G_{2}$. However, we recall that the longitudinal asymmetry in polarized
deep-inelastic lepton-nucleon scattering is sensitive to the nucleon singlet
axial-current matrix element, and use the value of $g_{A}^{0}$ extracted
from these experiments to estimate $G_{2}$:
\begin{equation}
g_{A}^{0}\;=\;0.3\pm0.1\;\rightarrow\;G_{2}\simeq2, \label{G2}%
\end{equation}
which is indeed small compared with $G_{10}$.

In order to predict $G_{\overline{10}}$ it is necessary to
know a new combination of $G_{0}$ and $G_{1}$. For this, we first seek
guidance from $\chi$SM calculations, which yield~\cite{Christov,DPP}:
\begin{equation}
G_{1}\;=\;(0.5\pm0.1)\times G_{0}.\label{G1overG0}%
\end{equation}
Using the central value in (\ref{G0G1}), we would find
\begin{equation}
G_{0}\simeq 17.9,\qquad G_{1}\simeq 8.9\label{G0andG1}%
\end{equation}
Inserting (\ref{G2}) and (\ref{G0andG1})
into (\ref{FoverD}), we find $F/D \simeq 0.59$, which should be compared
with the experimental value $0.56 \pm 0.02$. Moreover, the value
(\ref{G2}) worsens only slightly the leading-order
prediction (\ref{G2from_gpinn}) for $g_{\pi NN}$.

Inserting the values (\ref{G0andG1}, \ref{G2}) into the expression
(\ref{gKTN}), we find $G_{\overline{10}}=7.9$,
which is considerably smaller than either $g_{\pi
NN}$ or $g_{\pi\Delta N}$. However, this suppression is insufficient to
explain fully the narrow widths of the $\Theta^{+}$ and $\Xi^{--}$, as
suggested in~\cite{DPP}. For example, the total width of the $\Theta^{+}$,
which decays into $K N$, would be given by
\begin{equation}
\Gamma_{\Theta^+}\; = \; \frac{3G_{\overline{10}}^{2}}{8\pi
M_{\Theta^+}\,M_{N}} \frac{1}{5}\,p_{K}^{3} \; = \; 20.6~{\rm MeV}.
\end{equation}
Although this number is relatively small, it is considerably larger
than recent experimental estimates \cite{others}. For comparison, we recall
that the $\Theta^{+}$ decay width is formally of higher order in $1/N_{c}$
than that of the $\Delta$~\cite{Prasz11}, and that the CQM would suggest that
$G_{1}/G_{0}=4/5,G_{2}/G_{0}=2/5$, which would predict a strong suppression of
$\Gamma_{\Theta^{+}}$ and $\Gamma_{\Xi^{--}}$.

In view of the mixed success of the above calculation of baryon couplings
in the \raise 0.17em\hbox{$\chi$}SM, we explore the corrections due to
the initial assumption of universality in the coupling $g$ entering
(\ref{Lint}).

It was argued in~\cite{DPP} that the theoretical predictions for
the decay widths should be multiplied by the ratio $M^{\prime}/M$,
however their numerical values are consistent~\cite{Weigel} with
multiplying decuplet decays  by an inverse ratio~\footnote{The
authors \cite{DDPC} claim there was a misprint in~\cite{DPP},
where the ratio for the decuplet decays was inadvertently written
as $M^{\prime}/M$. Different opinions are presented 
in~\cite{Weigel,Jaffearith}.} $M/M^{\prime}$ and antidecuplet decays
by $M^{\prime}/M$. It is beyond the scope of the present paper to
discuss the origin of these corrections, here we try to examine
various approximations present in the literature, and this is just
one of the more important ones. The results below are displayed in
terms of multiplicative factors so it is easy to `undo' them, if
the reader would like to understand the impacts of different assumptions.

It is convenient to split these factors into $N_{c}$-dependent
and $m_s$ independent corrections
that are identical for the whole multiplet, and additional $m_{s}$
corrections that have to be calculated for each single decay
separately:
\begin{equation}
\frac{M}{M^{\prime}}=\frac{M_{10}}{M_{8}} \left(
\frac{M/M_{10}}{M^{\prime}/M_{8} } \right) = 1.2\times \left(
\frac{M/M_{10}}{M^{\prime}/M_{8} } \right) \equiv 1.2\times
R_{B\rightarrow B^{\prime}}^{(g)} \label{DPPfactors}
\end{equation}
where we have evaluated the $N_c$ dependent
correction using $M_{10}=1382.1$ MeV and
$M_{8}=1151.5$ MeV for the mean decuplet and
octet masses, respectively, and $R_{B\rightarrow B^{\prime}}^{(g)}$ is an
$m_{s}$-dependent
correction, which for the $\Delta\rightarrow N+\pi$ transition amounts to
$1.09$.
We see that inclusion of the factor $M_{10}/M_{8}$ reduces the value of
$G_{10}$ by $\sqrt{1.2}= 1.0954$, to
\begin{equation}
G_{10}=20.4
\label{G10204}
\end{equation}
resulting in
\begin{equation}
g_{\pi NN}\;=14.3+\frac{G_{2}}{20},\quad G_{\overline{10}}=7.2
\end{equation}
The width of $\Theta^{+}$ has to be modified now by the ratio $M_{8}%
/M_{\overline{10}}=0.66$ (for $M_{\overline{10}}=1754$~MeV), yielding
\begin{equation}
\Gamma_{\Theta^+}=\frac{3G_{\overline{10}}^{2}}{8\pi M_{\Theta^+}\,M_{N}%
}\frac{M_{8}}{M_{\overline{10}}}\frac{1}{5}\,p_{K}^{3}=11.1~{\rm MeV}
\end{equation}
which agrees with the original prediction of \cite{DPP}. The
excellent agreement in the second case is mainly due to the
suppression factor $M_{8}/M_{\overline{10}}$.

One may, alternatively, invert the logic and use the measured $\Theta^+$
width, which is
presumably smaller than $10$~MeV, to extract independently values of
$G_{0}$
and $G_{1}$. For this, we consider two extreme cases:
$\Gamma_{\Theta^+}=10$
and $1$~MeV. Then, without the $M_{8}/M_{\overline{10}}$ correction we
would obtain
\begin{eqnarray}
G_{\overline{10}} = 5.5 \quad &{\rm for~}& \Gamma_{\Theta^+}=10~{\rm MeV},
\nonumber \\
G_{\overline{10}} = 1.75 \quad &{\rm for~}& \Gamma_{\Theta^+}=1~{\rm MeV}
\end{eqnarray}
where we have chosen the positive sign in order to keep $G_{0}>G_{1}$.
Assuming $G_{2}=2$, we get two sets of solutions%
\begin{align}
G_{0}   =17.1,& \quad G_{1}=10.6 \quad {\rm for~} \Gamma_{\Theta^+}=10~{\rm MeV},
\nonumber \\
G_{0}   =15.8,& \quad G_{1}=13.1 \quad {\rm for~} \Gamma_{\Theta^+}=1~{\rm MeV}
\label{G0andG1_1}
\end{align}
The corresponding ratios $G_{1}/G_{0}$ lie somewhat outside the model
ranges quoted previously, but are still below unity.
We see that the freedom stemming from the fact that non-exotic
decays fix only one linear combination of $G_{0}$ and $G_{1}$ enables one
to accommodate a very narrow width of the
$\Theta^{+}$ without changing the prediction for $\Delta$.

We can now repeat the same for the expressions corrected by the multiplet
mass  ratios, and obtain
\begin{eqnarray}
G_{\overline{10}} = 6.82 \quad &{\rm for~}& \Gamma_{\Theta^+}=10~{\rm MeV},
\nonumber \\
G_{\overline{10}} = 2.16 \quad &{\rm for~}&
\Gamma_{\Theta^+}=1~{\rm MeV}.
\end{eqnarray}
Using $G_{10}=20.25$,
we get%
\begin{align}
G_{0}   =16.1, &\quad G_{1}=8.30 \quad {\rm for~} \Gamma_{\Theta^+}=10~{\rm MeV},
\nonumber\\
G_{0}   =14.6, &\quad G_{1}=11.4 \quad {\rm for~} \Gamma_{\Theta^+}=1~{\rm MeV}.
\label{G0andG1_2}
\end{align}
Note that
here the ratios $G_{1}/G_{0}$ are closer to the model range.

We remark that, irrespective of whether we correct the widths
by the $M_{R}/M_{R^{\prime}}$ ratio, or not,
\begin{eqnarray}
\Gamma_{\Xi^{--}_{\overline{10}}\rightarrow \Xi^- +\pi^-} &\sim&
1.3
\; \Gamma_{\Theta^+} \nonumber \\
\Gamma_{\Xi^{--}_{\overline{10}}\rightarrow \Sigma^- + K^-} &\sim&
0.8 \; \Gamma_{\Theta^+}
\end{eqnarray}
Hence, we predict total decay
widths for $\Theta$ and $\Xi^{--}$ which are similar to within a 
factor of 2. As we show below,
this relation is removed when we include symmetry-breaking terms.
Numerically the width of $\Xi_{\overline{10}}^{--}$, which was
recently estimated by the NA49 Collaboration to be below $18$~MeV
becomes $21$ and $2.1$~MeV for the two extreme cases
($\Gamma_{\Theta^+}=10$ and 1~MeV) discussed
above~\footnote{Another recent theoretical estimate puts the
$\Xi_{\overline{10}}$ width\,\,\,$\le 10$ MeV
\cite{Arndt:2003ga}.}.

For convenience in the subsequent analysis, we adopt the following
para\-metrization of the \chiSM\ model couplings:
\begin{equation}
G_1 \equiv \rho\ G_0, \;\;\;\;\; G_2 \equiv \epsilon \; G_0 = \left( \frac{9
(F/D)-5}{3 (F/D) +5} \right) (\rho+2) G_0
\label{rho}
\end{equation}
where the last equation follows from (\ref{FoverD}). The various fits
described above yield $\rho \lappeq 0.8$, as compared to the favoured
model range $\rho = 0.5 \pm 0.1$ (\ref{G1overG0}).

In order to check the sensitivity of the suppression mechanism
for $\overline{10}$ decays to the numerical values of couplings
$G_i$, we plot in Fig. 1 the ratios $(G_{\overline{10}}/G_{10})^2$
for different values of  $F/D$ as functions of the parameter $\rho$.
We see that $\overline{10}$ decays are suppressed with respect
to $10$ for a wide range of $\rho$. Further suppression,
if one follows the logic of~\cite{DPP},
may be provided, as discussed above, by the rescaling of the decay
widths by the average multiplet mass ratios, and by the $m_s$ corrections,
as discussed in the following section.
In the numerical evaluations below, we vary $\rho$ from 0.2 to 0.8 while
fixing $F/D=0.59$.

\begin{figure}[h]
\centerline{\epsfysize=3.5in\epsffile{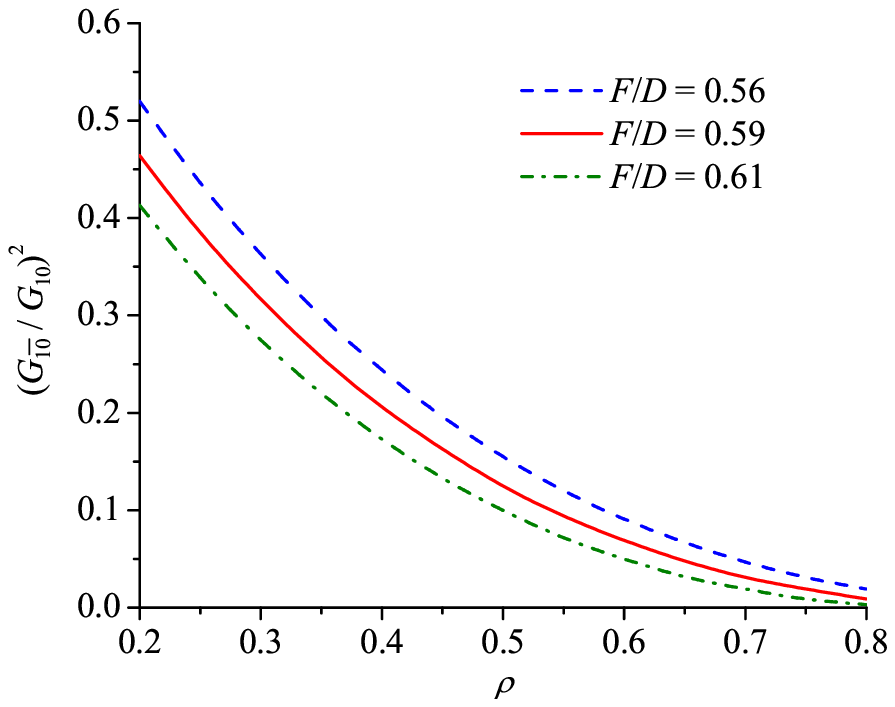}} \vskip4pt
{\small Fig. 1. {\it
Values of the ratios $(G_{\overline{10}}/G_{10})^2$
for different values of $F/D$, as functions of the parameter $\rho =
G_1/G_0$.
}}
\end{figure}

\subsection{SU(3) Symmetry Breaking Effects}

We now calculate the SU(3)-breaking corrections due to the baryon
representation mixing and the non-universality of the $g$ coupling
discussed above, as an aid to assessing the
SU(3)-breaking uncertainties in the analysis of Sect.~\ref{symm}.
There is some
ambiguity in the way these corrections are treated. As already discussed,
by using experimental
values for the masses $M$ and $M^{\prime}$ in (\ref{Gamdef1}), as well as
for $p_{\varphi}$, we \textit{implicitly} sum an infinite series of
$m_{s}/\Lambda_{QCD}$ corrections. In the following, however, we compute
the square of $\mathcal{M}$ up to terms linear in $m_{s}$ stemming
from the representation mixing due to the mass
splitting hamiltonian $H^{\prime}$. We recall that there are also
corrections to the decay operator $O_{\varphi}^{(8)}$, which -- as
explained above -- are ignored in the following. We shall also include
residual $m_{s}$ corrections coming from the ratios
$(M/M_{10})/(M^{\prime}/M_{8})$ for decuplet decays and
$(M^{\prime }/M_{8})/(M/M_{\overline{10}})$ for antidecuplet decays,
respectively.

Decuplet baryons can decay only to octet baryons, and we have%
\begin{align}
\left\langle B_{8}^{\prime}\right|  \hat{O}_{\varphi}^{(8)}\left|
B_{10}\right\rangle  &  =\left\langle 8_{1/2},B^{\prime}\right|
\hat{O}_{\varphi}^{(8)}\left|  10_{3/2},B\right\rangle \nonumber\\
&  +a_{27}^{B}\left\langle 8_{1/2},B^{\prime}\right|  \hat{O}_{\varphi}%
^{(8)}\left|  27_{3/2},B\right\rangle +c_{27}^{B^{\prime}}\left\langle
27_{1/2},B^{\prime}\right|  \hat{O}_{\varphi}^{(8)}\left|  10_{3/2}%
,B\right\rangle ,
\end{align}
where we include mixings with the 27 multiplets of baryons, which were
neglected in~\cite{DPP,DP9}. We then introduce \cite{MichalPREP}
\begin{equation}
G_{10}\equiv G_{0}+\frac{1}{2}G_{1},\;G_{27}\equiv G_{0}-\frac{1}{2}%
G_{1},\;G_{27}^{\prime}\equiv G_{0}-2G_{1}. \label{Gs}%
\end{equation}
In terms of these combinations, the baryon representation mixing discussed
earlier yields \cite{MichalPREP}
\begin{equation}
\overline{\mathcal{M}_{\Delta}^{2}}=\frac{3}{5} G_{10} \left[
G_{10} +
\frac{10}{3}a_{27}\,G_{27}+\frac{2}{3}c_{27}\,G_{27}^{\prime}\right]
p_{\varphi}^{2}%
\end{equation}
for the squared matrix element to first order in $m_{s}/\Lambda_{QCD}$.

Using the previous numbers for the mixing coefficients, the ratio of the new
expression for $\Delta$ decay to the old one is
\[
R_{\Delta\rightarrow N}^{(mix)}=1+\frac{0.499\,G_{27}+0.042 \,G_{27}%
^{^{\prime}}}{G_{10}}%
\]
It can be seen from Figure 2 that $R_{\Delta\rightarrow
N}^{(mix)}$ is quite insensitive to the value of parameter $\rho$.
For the specific value $\rho=1/2$, we find

\begin{equation}
R_{\Delta\rightarrow N}=R_{\Delta\rightarrow
N}^{(g)}R_{\Delta\rightarrow N}^{(mix)} \; = \; 1.09 \times 1.3 \;
= \;1.42
\end{equation}
in which case the value of $G_{10}$ extracted in (\ref{G10204})
should be reduced by a factor 1/1.19, to
\begin{equation}
G_{10} = 17.1
\label{G10with3b}
\end{equation}
when we include the SU(3) breaking due to representation mixing.
On the other hand, if we neglect factors of mass ratios from
(\ref{DPPfactors}), then we should reduce the coupling of
(\ref{G0G1}) by a factor 1/1.14, to
\begin{equation}
G_{10} = 19.65. \label{G10with3c}
\end{equation}
In either case, the $G_{10}$ coupling is reduced and, as we
see below, the prediction for $g_{\pi NN}$ is improved.

\begin{figure}[h]
\centerline{\epsfysize=3.5in\epsffile{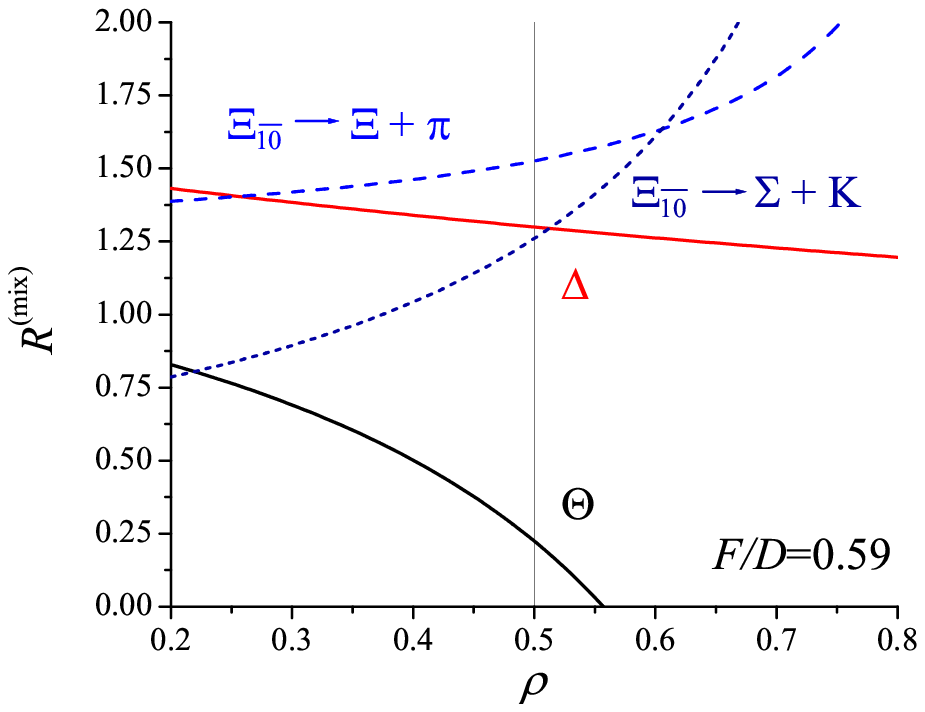}} \vskip4pt
{\small Fig. 2. {\it The correction factors $R^{(mix)}$ due to
SU(3)-breaking representation mixing for the decays discussed in
the text, as functions of the parameter $\rho \equiv G_1 / G_0$.}}
\end{figure}

Finally, we calculate the corresponding SU(3) corrections to the
$\pi$-nucleon coupling constant due to representation mixing,
which can be obtained from the formula
\begin{equation}
g_{\pi NN}=\left| \left\langle p_{8}\right|  \hat{O}_{\pi^{0}}^{(8)}\left|
p_{8}\right\rangle\right| \frac{1}{p_{\pi}} \label{gpipi}%
\end{equation}
where we work in the frame $\vec{p}_{\pi}=(0,0,p_{\pi})$.
This gives \cite{MichalPREP}
\begin{equation}
g_{\pi NN}=\frac{7}{10}\left[  G_{0}+\frac{1}{2}G_{1}+\frac{1}{14}%
G_{2}\right]
+c_{\overline{10}}G_{\overline{10}}+\frac{2}{15}c_{27}H_{27},
\label{firstgpinn}
\end{equation}
where we have introduced
\begin{equation}
G_{\overline{10}} \equiv G_{0}-G_{1}-\frac{1}{2}G_{2},\;\;\;\;H_{27}\equiv
G_{0}-2G_{1}+\frac{3}{2}G_{2}.
\label{G10barH27}
\end{equation}
The expression (\ref{firstgpinn}) may be written in the form
\begin{equation}
g_{\pi NN}=G_{10} \times \left( {{0.796 + 0.245 \rho + 0.019 \epsilon}
\over {1 + 0.5 \rho}} \right),
\label{finalgpinn}
\end{equation}
which yields \ $g_{\pi NN} = 13.2$ to 12.2 \ for $G_{10} = 17.1$,
as found in (\ref{G10with3b}) after including representation
mixing, and $\rho = 0.2$ to 0.8. This result obtained including
representation mixing compares better with the experimental range
$g_{\pi NN}=13.3\pm 0.1$~\cite{Ericson} or $13.13\pm
0.07$~\cite{Bugg:2003bj} than did the leading-order prediction
(\ref{G2from_gpinn}), and leaves open the possibility that a more
complete calculation of $m_{s}/\Lambda_{QCD}$ and ${\cal
O}(1/N_c)$ effects - including those discussed in~\cite{Dorey} -
might remove the discrepancy completely. We plot in Fig.~3 the
$\pi$-nucleon coupling $g_{\pi NN}$ as a function of the parameter
$\rho \equiv G_1 / G_0$ for $F/D = 0.59$ and four different values
of $G_{10}$ discussed in the text (\ref{G0G1}, \ref{G10204},
\ref{G10with3b},\ref{G10with3c}). In general, this example warns
us that, to the accuracy they are currently made, \raise
0.17em\hbox{$\chi$}SM calculations of couplings are subject to
uncertainties of $\mathcal{O}(20)$\%.

\begin{figure}[h]
\centerline{\epsfysize=3.5in\epsffile{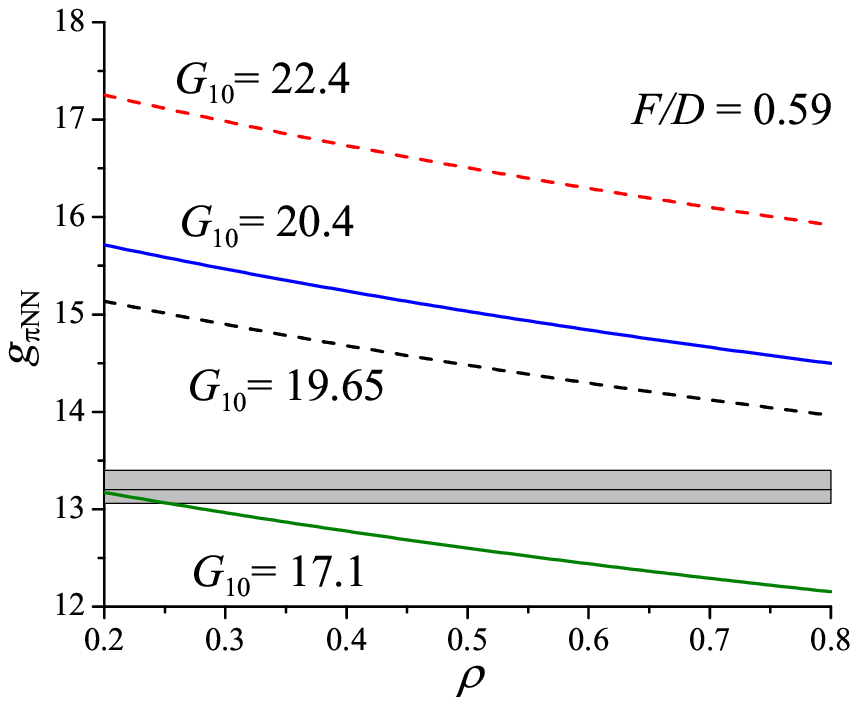}} \vskip4pt
{\small Fig. 3. {\it The $\pi$-nucleon coupling $g_{\pi NN}$ as a
function of the parameter $\rho \equiv G_1 / G_0$ for $F/D = 0.59$
and four different values of $G_{10}$ discussed in the text
(\ref{G0G1}, \ref{G10204}, \ref{G10with3b}, \ref{G10with3c}). The
upper shaded band corresponds to the experimental evaluation
of~\cite{Ericson}, and the lower shaded band corresponds to the
range in~\cite{Bugg:2003bj}. }}
\end{figure}

In the absence of SU(3)-symmetry breaking, antidecuplet baryons can decay
directly only to octet baryons. Including first-order $m_{s}/\Lambda_{QCD}$
effects, we find:%
\begin{align}
\left\langle B_{8}^{\prime}\right|  \hat{O}_{\varphi}^{(8)}\left|
B_{\overline{10}}\right\rangle  &  =\left\langle 8_{1/2},B^{\prime}\right|
\hat{O}_{\varphi}^{(8)}\left|  \overline{10}_{1/2},B\right\rangle \nonumber\\
&  +d_{8}^{B}\left\langle 8_{1/2},B^{\prime}\right|  \hat{O}_{\varphi}%
^{(8)}\left|  8_{1/2},B\right\rangle +d_{27}^{B}\left\langle 8_{1/2}%
,B^{\prime}\right|  \hat{O}_{\varphi}^{(8)}\left|  27_{1/2},B\right\rangle
\nonumber\\
&  +c_{\overline{10}}^{B^{\prime}}\left\langle \overline{10}_{1/2},B^{\prime
}\right|  \hat{O}_{\varphi}^{(8)}\left|  \overline{10}_{1/2},B\right\rangle
+c_{27}^{B^{\prime}}\left\langle 27_{1/2},B^{\prime}\right|  \hat{O}_{\varphi
}^{(8)}\left|  \overline{10}_{1/2},B\right\rangle , \label{fullmel}
\end{align}
where only the term proportional to $c_{\overline{10}}^{B^{\prime}}$ and
$d_{27}^{B}$ were taken into account previously~\cite{DPP,DP9}. In the
presence of SU(3) breaking, decays of antidecuplet baryons into decuplet
baryons also become possible, via a matrix element
\begin{eqnarray}
\left\langle B_{10}^{\prime}\right|  \hat{O}_{\pi^{-}}^{(8)}\left|
B_{\overline{10}}\right\rangle = d_{27}^{B}\left\langle 10_{3/2},B^{\prime
}\right|  \hat{O}_{\varphi}^{(8)}\left|  27_{1/2},B\right\rangle
+a_{27}^{B^{\prime}}\left\langle 27_{3/2},B^{\prime}\right|  \hat{O}_{\varphi
}^{(8)}\left|  \overline{10}_{1/2},B\right\rangle .
\kern-3em\strut
\nonumber \\
\label{10barto10}
\end{eqnarray}
whose magnitude we discuss later.

For the discussion of these decays, we introduce the following constants
\cite{MichalPREP}%
\begin{align}
H_{\overline{10}}  &  \equiv G_{0}-\frac{5}{2}G_{1}+\frac{1}{2}G_{2}%
,\;H_{\overline{10}}^{\prime} \equiv G_{0}+\frac{11}{14}G_{1}+\frac{3}%
{14}G_{2},\nonumber\\
H_{8}  &  \equiv G_{0}+\frac{1}{2}G_{1}-\frac{1}{2}G_{2},\;\,H_{8}^{\prime
}\equiv G_{0}+\frac{1}{2}G_{1}-\frac{1}{6}G_{2}.
\end{align}
In terms of these, we find for the average of the $\Theta^{+}\rightarrow
p+K^{0}$ and $\Theta^{+}\rightarrow n+K^{+}$ decays:

\begin{equation}
\overline{\mathcal{M}^{2}}_{\Theta^+\rightarrow N+K}=\frac{3}{10}
G_{\overline{10}} \left[
G_{\overline{10}}+\frac{5}{2}c_{\overline{10}}\,H_{\overline{10}}-\frac{7}%
{2}c_{27}H_{\overline{10}}^{\prime}\right] \times p^{2}.%
\label{breaktheta}
\end{equation}
This formula resembles that given in~\cite{DPP}, but there are some differences:

\begin{itemize}
\item The squared decay matrix element is not just a function of
$G_{\overline{10}}$. In fact, when $|G_{\overline{10}}|$ is comparable
to the SU(3)-breaking corrections, one should use the full
quadratic expression for
$\overline{\mathcal{M}^{2}}_{\Theta^+\rightarrow N+K}$, rather
than the linear form (\ref{breaktheta}). Even if
$G_{\overline{10}} = 0$, $\Theta^+$
could still decay through the mixing terms, contrary to the impression given
by eq.~(56) of~\cite{DPP} (see however \cite{Arndt:2003ga}).

\item There is an additional term due to mixing with the $27$ representation,
which is not small and was not included in~\cite{DPP}.

\item The squaring of the matrix element introduces a factor of $2$ which is
not apparent in~\cite{DPP}.
This has been corrected in \cite{Arndt:2003ga}.
\end{itemize}

The corresponding squared amplitude for the $\Xi_{\overline{10}}^{--}\rightarrow
\Xi^{-}+\pi^{-}$ decay reported by NA49 has the form:%
\begin{equation}
\overline{\mathcal{M}^{2}}_{\Xi_{\overline{10}}^{--}\rightarrow\Xi^{-}+\pi^{-}%
}=\frac{3}{10} G_{\overline{10}} \left[
G_{\overline{10}}+\frac{7}{3}c_{27}\,H_{\overline{10}%
}^{\prime}+\frac{2}{3}d_{27}H_{27}\right] \times p^{2}.
\end{equation}
Another possible decay of this state is $\Xi_{\overline{10}}^{--}%
\rightarrow\Sigma^{-}+K^{-}$, whose squared matrix element is given by
\begin{equation}
\overline{\mathcal{M}^{2}}_{\Xi_{\overline{10}}^{--}\rightarrow\Sigma^{-}+K^{-}%
}=\frac{3}{10} G_{\overline{10}} \left[
G_{\overline{10}}-\frac{5}{2}c_{\overline{10}%
}\,H_{\overline{10}}+\frac{7}{6}c_{27}\,H_{\overline{10}}^{\prime}%
-\frac{2}{3}d_{27}\,H_{27}\right] \times p^{2}.
\end{equation}
These examples exhibit explicitly that the SU(3)-breaking corrections to
${\overline{10}}$ decays are \textit{not} universal.

Let us define the correction factor coming from the representation
mixing:
\begin{equation}
R^{(mix)}_{\Theta^+\rightarrow N+K}=1 +\frac{0.22\
H_{\overline{10}}
      -0.22\ H^{\prime}_{\overline{10}}}{G_{\overline{10}}}.
    \end{equation}
In Fig.~2 we plot $R^{(mix)}_{\Theta^+\rightarrow N+K}$ as a function of
parameter $\rho$. It can be seen that it is rather sensitive to value
of $\rho$, yielding for $\rho=1/2$
\begin{equation}
R^{(mix)}_{\Theta^+\rightarrow N+K}= 0.2.
\end{equation}
The correction from the non-universality of the $g$
coupling is
\begin{equation}
R^{(g)}_{\Theta^+\rightarrow N+K}=\frac{M_N/M_8}{M_{\Theta^+}/M_{\overline{10}}}
=0.93.
\end{equation}
These two corrections act in a similar way, tending to suppress
the decay rate of $\Theta^+$ by a further factor of $\sim 0.25$,
reinforcing the \chiSM\ prediction that the $\Theta^+$ should be very
narrow, and emphasizing that the SU(3)-breaking corrections are
potentially very significant in this case.

In the case of the $\Xi$ decays, we have
\begin{eqnarray}
R^{(mix)}_{\Xi^{--} \to\Xi^{-} + \pi^{-}} &=& 1+\frac{0.15\
H^{\prime}_{\overline{10}}+0.16\ H_{27}}{G_{\overline{10}}},
\nonumber \\
R^{(mix)}_{\Xi^{--} \to\Sigma^{-} + K^{-}} &=& 1+\frac{-0.22\
H_{\overline{10}}+0.07\ H^{\prime}_{\overline{10}}-0.16\
H_{27}}{G_{\overline{10}}}.
\end{eqnarray}
We see from Fig.~2 that $R^{(mix)}_{\Xi^{--} \to\Xi^{-} +
\pi^{-}}$ is a slowly-varying function of $\rho$, while
$R^{(mix)}_{\Xi^{--} \to\Sigma^{-} + K^{-}}$ is close to 1 in the
vicinity of $\rho=1/2$. Numerically, for $\rho=1/2$ we obtain:
\begin{eqnarray}
R^{(mix)}_{\Xi^{--} \to\Xi^{-} + \pi^{-}}&=&1.535, \nonumber \\
R^{(mix)}_{\Xi^{--} \to\Sigma^{-} + K^{-}}&=&1.269.
\end{eqnarray}
The $g$ correction reads in this case
\begin{eqnarray}
R^{(g)}_{\Xi^{--} \to\Xi^{-}+ \pi^{-}}&=&
\frac{M_{\Xi}/M_8}{M_{\Xi_{\overline{10}}}/M_{\overline{10}}}=1.08,
\nonumber \\
R^{(g)}_{\Xi^{--} \to\Sigma^{-} + K^{-}} &=&
\frac{M_{\Sigma}/M_8}{M_{\Xi_{\overline{10}}}/M_{\overline{10}}}=0.98.
\end{eqnarray}
Despite small suppression in the last case, the $m_s$ corrections
tend to increase the width of $\Xi_{\overline{10}}$, reinforcing
the message that the corrections to ${\overline{10}}$ decays are
not universal.

Finally, we consider the decay
$\Xi_{\overline{10}}^{-}\rightarrow\Xi^{\ast0}+\pi^{-}$,
preliminary evidence for which
was recently mentioned by NA49 \cite{KadijaJLab}.
Since this decay is not
allowed in the SU(3) symmetry limit, it can only go via admixtures of $27$
multiplets in the $\overline{10}$ and/or $10$,
as given in (\ref{10barto10}). Calculating the relevant matrix element,
we get \cite{MichalPREP}
\begin{equation}
\overline{\mathcal{M}^{2}}_{\Xi_{\overline{10}}^{-}\rightarrow\Xi^{\ast0}+\pi^-%
}=\frac{1}{162}\left[  d_{27}\left(  G_{0}-2G_{1}\right)  +a_{27}\left(
G_{0}+G_{1}\right)  \right]  ^{2}p^{2}.
\end{equation}
This matrix element is extremely small, approximately two orders of
magnitudes smaller than the one for $\Theta^+$ decay~\footnote{Note
that the meson momenta $p$ are identical for both decays, to within 2
MeV.} (\ref{breaktheta}). Furthermore, the masses in the denominator of
(\ref{Gamdef1}) give another factor of $1/2$, yielding the decay rate
\begin{equation}
\Gamma_{\Xi_{\overline{10}}^{-}\rightarrow\Xi^{\ast0}+\pi^{-}} \sim
\left( \frac{1}{200} \div \frac{1}{100} \right) \Gamma_{\Theta^+}.
\end{equation}
Therefore this mixing mechanism is unlikely to be
the explanation of the preliminary evidence reported by NA49.

Within the CQM, an
interpretation of this decay as due to the decay of an isodoublet $\Xi$
state within an
octet of pentaquarks, which is degenerate with the $\Xi$ in the
$\overline{10}$, was recently proposed~\cite{JW2}. There is no additional
rotational octet excitation in the
\chiSM, and it was therefore argued in~\cite{JW2} that the confirmation of
this decay would be a challenge for the \chiSM. However, we remark that
octets are expected as vibrational excitations in the \chiSM, but with
properties that are very difficult to estimate. Nevertheless, $1/N_c$
arguments suggest that these vibrational excitations should have masses
comparable to the exotic rotational excitations discussed above. An
alternative explanation of the $\Xi_{\overline{10}}^{-} \rightarrow
\Xi^{\ast0} + \pi^{-}$ decay, offered in the next Section, is that the
$\Xi^{-}$ state reportedly observed is a member of the $(27, {3 \over 2})$
that might be almost degenerate with that in the ${\overline{10}}$.

Let us briefly summarize the findings of this Section. First, we have
shown that the $m_s / \Lambda$ corrections are not universal. Secondly,
they are rather large and in some cases, such as the $\Theta^+$ decay
rate, sensitive to the parameter $\rho=G_1/G_0$ which we have varied
between 0.2 and 0.8. One should not be surprised that these corrections
are large, since the leading term is small and vanishes exactly in the
quark model limit of the \chiSM~\footnote{Strictly speaking, the
enhancement factors $R^{(mix)}$ would diverge in this limit and lose
physical meaning.}, whereas no other matrix element vanishes in this
limit. This means that some antidecuplet decays may be controlled
primarily by
representation mixing. Thirdly, we have calculated the decay width
of $\Xi_{3/2}$ to $\Xi^{\ast}(1530)$ which can only go through
the admixture of 27 and found out that it was 2 orders of magnitude
smaller than the width of $\Theta^+$.

\section{Predictions for the Masses of Other Exotic Baryons}

As already mentioned, the SU(3) \chiSM\ predicts a tower of
heavier and more exotic baryons, of which the lightest is expected
to be a $27$ representation with $J^P = {3 \over 2}^+$. Several
numerical estimates have been made for the masses of the exotic
$\Theta_1$, $\Sigma_2$ and $\Omega_1$ baryons in these
multiplets~\cite{Kopel,other27}, where the symbols specify the
strangeness (hypercharge) and the subscripts specify the isospins
of these states. In light of the previous analysis, using the
masses of the $\Theta^+$ and $\Xi_{\overline{10}}$ as inputs, we
now refine these predictions.

We recall that the splittings between the centres of the lowest-lying
27-plet, octet and decuplet baryons are given in the \chiSM\ by
\begin{equation}
\Delta M_{(27, {3 \over 2}) - (10, {3 \over 2})} \; = \; {1 \over I_2}, \;
\;
\Delta M_{(27, {1 \over 2}) - (8, {1 \over 2})} \; = \; {5 \over 2 I_2},
\label{DeltaM27}
\end{equation}
%
%
the chiral-symmetry breaking mass corrections within the $(27, {3 \over
2})$ multiplet are
\begin{eqnarray}
\Theta_1 \; &:& \; + {1 \over 7} \alpha + 2 \beta - {5 \over 14} \gamma, \\
\Sigma_2 \; &:& \; + {5 \over 56} \alpha - {25 \over 112} \gamma, \\
\Omega_1 \; &:& \; - {13 \over 56} \alpha - 2 \beta + {65 \over 112} \gamma,
\label{273masses}
\end{eqnarray}
where the subscript denotes the isospin of a given baryon in the 27-plet.
Using the values of $I_2, \alpha, \beta$ and $\gamma$ extracted previously
(\ref{exptnumbers}) from the observed $\Theta^+$ and $\Xi_{\overline{10}}$
 masses, we
estimate for the exotic baryons in the $(27, {3 \over 2})$
multiplet:
\begin{equation}
\begin{array}{ccc}
(27, {3 \over 2}): & \; m_{\Theta_1} \; = \; 1597~{\rm MeV},&\; m_{\Sigma_2}
\; = \; 1695~{\rm MeV}, \\
~~& m_{\Xi_{3/2}}=1876~{\rm MeV}, &\; m_{\Omega_1} \; = \; 2057~{\rm MeV}.
\end{array}
\label{273numbers}
\end{equation}
We note that the $\Xi_{3/2}$ in the $(27, {3 \over 2})$ is almost
degenerate
with $\Xi_{3/2}$ in $\overline{10}$.
As discussed in the previous Section,
this might be relevant to the preliminary
evidence of a state at 1860 MeV decaying into
$\Xi(1530)^0 + \pi$~\cite{KadijaJLab}.
Such a decay is not allowed for the
$\Xi_{3/2}$ in the $\overline{10}$, since $\overline{10} \notin
10\times 8$,
but it would be allowed for a $\Xi_{3/2}$ in the 27, since $27 \in
10\times 8$.

\begin{figure}[h]
\centerline{\epsfysize=4.1in\epsffile{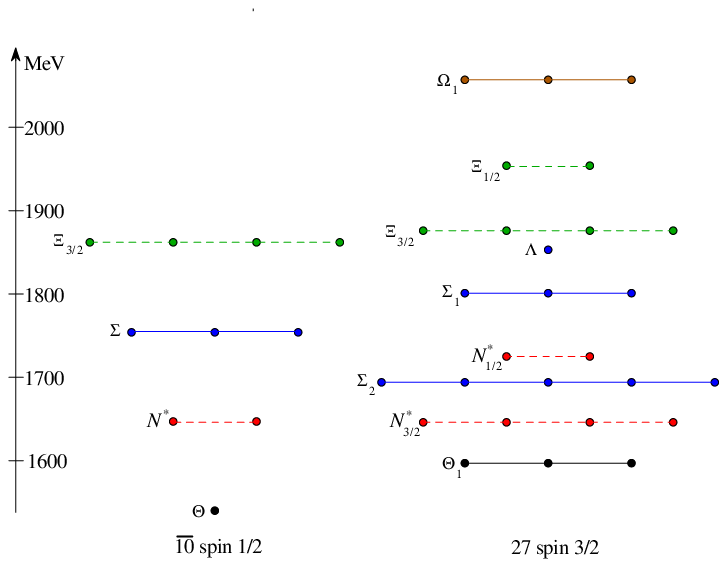}} \vskip4pt
{\small Fig. 4. {\it The spectra of exotic baryons found
at first order in SU(3) symmetry breaking, using parameters
fitted from the $\Theta^+$ and $\Xi_{\overline{10}}$ masses.
The $({\overline {10}}, {1 \over 2}^+)$ spectrum is shown on the left, and
the $(27,{3 \over 2}^+)$  spectrum on the right.}}
\end{figure}

The spectra of the exotic baryons found at first order in $SU(3)$ symmetry
breaking in the $\overline{10}$ and $(27,{3\over2}^+)$ representations are
shown in Fig. 4.

It should be emphasized that the $1 / N_c$ expansion used in the
\chiSM\ approach becomes less reliable for heavier baryons, so
these numerical predictions should be treated as only approximate.
However, we confirm previous suggestions~\cite{Kopel,other27} that
there may be an isospin triplet of $S = - 1$ $\Theta_1$ baryons
weighing barely 60~MeV more than the $\Theta^+$, and the presence
of a low-lying $I{=}2\,$ $\Sigma$ multiplet is also suggested.
These would both have $J^P = {3 \over 2}^+$, with the
corresponding $(27, {1 \over 2}^+)$ being significantly heavier.
If found, these exotic $27$ baryons would provide further
encouragement for the \chiSM\ approach.

For comparison, a recent detailed study~\cite{Bijker} of exotic baryon
spectroscopy in the CQM suggests the existence of a $({\overline {10}}, {3
\over 2}^+)$ excitation of the $\Theta^+$ with a mass within about 100~MeV
of the $\Theta^+$ (see also~\cite{DC}), a slightly heavier $\Theta_1$
state in the $(27, {1 \over 2}^+)$ and a rather heavier $\Theta_1$ state
in the $(27, {3 \over 2}^+)$. In this approach, the exotic baryons with $Y
< 2$ are significantly lighter than in our \chiSM\ estimates above: in
particular, the $\Xi$ state in the $({\overline {10}}, {1 \over
2}^+)$ is considerably lighter than was recently reported~\cite{Xi}.

It is interesting to exhibit explicitly the mass difference of the lightest
members of the $(\overline{10},\frac{1}{2})$ and $(27,\frac{3}{2})$
multiplets:
\begin{equation}
\Delta_{\Theta}=M_{\Theta_{27}}-M_{\Theta^+_{\overline{10}}}=\frac{1}{2}
\left( 3\frac{1}{I_1}-\frac{1}{I_2} \right)
-\frac{1}{56}\left( 13 \gamma+ 6 \alpha \right).
\end{equation}
We see that, for the set of parameters (\ref{exptnumbers}), partial
cancellations occur in each bracket, yielding $\Delta_{\Theta}=63$~MeV.
The lowest isospin triplet in the $(27,\frac{3}{2})$ multiplet is only
slightly heavier than the $\Theta^+$. A similar cancellation occurs
for $\Xi$ states:
\begin{equation}
\Delta_{\Xi}=M_{\Xi_{27}}-M_{\Xi_{\overline{10}}}
=\frac{1}{2}
\left( 3\frac{1}{I_1}-\frac{1}{I_2} \right)
+\frac{1}{112}\left( 13 \gamma+ 6 \alpha \right).
\end{equation}
which yields $\Delta_{\Xi}=18.7$~MeV for the set of parameters
(\ref{exptnumbers}).

Although this looks like an accidental cancellation, it is actually quite
robust, and would persist even if we did not assume that the mass of the
$\Xi_{\overline{10}}$ is 1860~MeV. This is illustrated in Fig. 5, where
we plot the $\overline{10}$ spectrum, together with the (27, $\frac{3}{2}$)
states $\Theta_1$ and $\Xi_{3/2}$ (dashed lines) as functions of the
$\pi$-nucleon sigma term $\SigmapN$. In making this plot, we have
taken as inputs only
the masses of the non-exotic states and of the $\Theta^+$, in
order to determine $\alpha$, $\beta$, $\gamma$ and $I_2$, but have
not used the mass of the $\Xi_{\overline{10}}$.
We see that the
lowest (27, $\frac{3}{2}$) state $\Theta_1$ is only a few tens of
MeV above $\Theta^+$, and that the $\Xi$ states are almost
degenerate, for a large range of $\SigmapN$.

\begin{figure}[h]
\centerline{\epsfysize=3.5in\epsffile{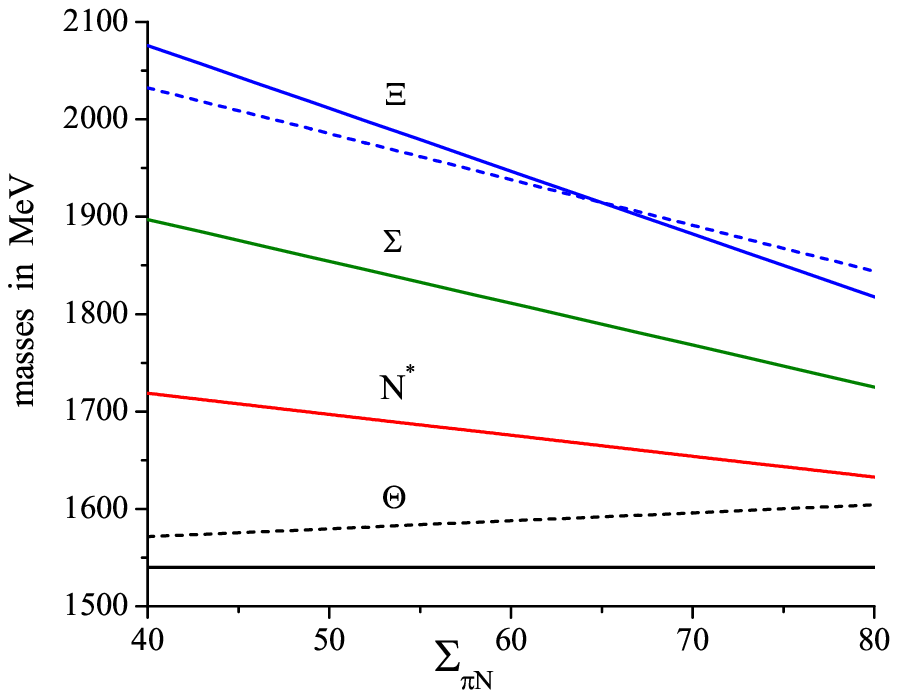}} \vskip2pt
{\small Fig. 5. {\it The spectra of $(\overline{10}, {1 \over 2})$ baryons
(solid lines) together with the masses of the $\Theta_1$ and $\Xi_{3/2}$
in the (27, $\frac{3}{2}$) (dashed lines) as functions of $\Sigma_{\pi
N}$, using parameters fitted from the masses of the $\Theta^+$ and
non-exotic states.}}
\end{figure}

One should therefore consider the possibility that NA49 has already
seen the $\Xi_{27}$ state decaying
to $\Xi^{\ast}(1530)$. In order to test this hypothesis, let
us calculate the decay width:
\begin{equation}
\Gamma_{B_{27}\rightarrow
B_{10}^{\prime}\varphi}=\frac{1}{8\pi}\frac{F_{27}%
^{2}}{MM^{\prime}}\frac{25}{72}\left(
\begin{array}
[c]{cc}%
8 & 10\\
\varphi & B^{\prime}%
\end{array}
\right.  \left|
\begin{array}
[c]{c}%
27\\
B
\end{array}
\right)  ^{2}p_{\varphi}^{3} \label{F27to10}
\end{equation}
where \cite{MichalPREP}
\begin{equation}
F_{27}=G_0-\frac{1}{2} G_1-\frac{3}{2} G_2.
\end{equation}
Let us note that, similarly to $G_{\overline{10}}$,
$F_{27}$ vanishes in the CQM limit (i.e., for
$G_1/G_0=4/5$ and $G_2/G_0=2/5$), and so we expect the decay
width (\ref{F27to10}) to be small. Indeed, for the values of
$G_{1,2}$ given in (\ref{G0andG1_1},\ref{G0andG1_2}) we
get $F_{27}=6$ to $9$, which is still bigger than
$G_{\overline{10}}$ but smaller than $G_{10}$. Moreover,
for the decay $\Xi^-(1560)\rightarrow\Xi^{\ast}(1530)+\pi^0$
there is another suppression factor, namely the square of the
SU(3) Clebsch-Gordan coefficient entering (\ref{F27to10}),
which is $1/6$. Altogether the width is of the order of 1 MeV:
\begin{equation}
\Gamma_{\Xi^-_{27} \rightarrow\Xi^{\ast\;0}_{10}+\pi^-} \sim 0.6
\div 1.5~{\rm MeV},
\label{XitoXi}
\end{equation}
depending on the value of $F_{27}$ and the mass
of $\Xi^-_{27}$.

For $27_{3/2}\rightarrow 8_{1/2}$ we obtain
(not summed or averaged over isospin):%
\begin{equation}
\Gamma_{B_{27}\rightarrow
B_{8}^{\prime}\varphi}=\frac{1}{8\pi}\frac{G_{27}%
^{2}}{MM^{\prime}}\frac{4}{9}\left(
\begin{array}
[c]{cc}%
8 & 8\\
\varphi & B^{\prime}%
\end{array}
\right.  \left|
\begin{array}
[c]{c}%
27\\
B
\end{array}
\right)  ^{2}p_{\varphi}^{3}\label{G27to8}%
\end{equation}
where%
\begin{equation}
G_{27}=G_{0}-\frac{1}{2}G_{1}=G_{10}-G_{1}.
\end{equation}
For the values of $G_{1,2}$ given in (\ref{G0andG1_1},\ref{G0andG1_2})
we get $G_{27}=9$ to $12$, yielding rather large 27-plet widths.
In the case of the $\Theta_{27}^{++}$, the Clebsch-Gordan
coefficient in (\ref{G27to8}) is unity, and we get
\begin{equation}
\Gamma_{\Theta_{27}^{++}\rightarrow p+K^{+}}\sim 37 \div 66~{\rm MeV}.
\label{ThetaXXVIIwidth}
\end{equation}
This is rather larger
than the width of the $\Theta^+_{\overline{10}}$.
Moreover, since this is a decay from spin $3/2$ to spin $1/2$,
the logic of~\cite{DPP} would imply a non-universality factor
$M/M^{\prime}$ that would increase the 27 widths even further.
Searches for
$I=1$ `partners' of the $\Theta^+_{\overline{10}}$ need to take this into
account, together with the negative results of previous experimental
searches~\cite{Hyslop:cs}.

Similarly, we obtain $41\div77$ MeV for the total width of
$\Xi(27,{3\over2})^-$, implying a very small branching ratio
$\lsim 0.02$ for the decay into $\Xi(1530)+ \pi^-$, shown
in \eqref{XitoXi}.
This poses two challenges for
the interpretation of the preliminary NA49 data \cite{KadijaJLab}
as decay of a $\Xi^-(27,{3\over2})$: one is that
the total production rate of the $\Xi^-(27,{3\over2})$
would need to be larger
by a factor 50 or so to compensate for the small branching ratio, and the
other is that the natural width would probably exceed the NA49 limit.

\section{Summary}

We have examined carefully the predictions of the \chiSM\ for the masses
and widths of the exotic baryons $\Theta^+$ and $\Xi_{\overline{10}}$.
It was a
non-trivial success for the \chiSM\ to have predicted the existence of
such relatively light exotic states~\cite{DPP}, candidate members of a
novel ${\overline {10}}$ multiplet of baryons~\cite{antidec}. The old
complaint that the \chiSM\ predicts unobserved exotic particles has been
refuted. The CQM did not predict such states, although it may accommodate
them. A key untested prediction of the \chiSM\ is that the $\Theta^+$ and
$\Xi_{\overline{10}}$ should have $J^P = {1 \over 2}^+$.
Some versions of the CQM
suggest instead $J^P = {1 \over 2}^-$, but $J^P = {1 \over 2}^+$ can be
accommodated in variants of the CQM with strongly-bound
diquarks~\cite{KL,JW}.

Dynamical calculations of soliton moments of
inertia~\cite{Prasz,moments} and a realistic assessment of our
knowledge of chiral symmetry breaking contributions to baryon
masses~\cite{Sigma} could have been used to predict ranges for
their masses that include the observed values, but with
uncertainties $\sim 200$~MeV. The remarkable prediction
of~\cite{DPP}, although somewhat fortuitous, exhibits an important
feature of the soliton models, namely the fact that exotic states
are much lighter than naive expectations of the quark model, which
would predict the lightest strange pentaquark to weight of the
order of 1700 MeV. This \chiSM\ is an inevitable consequence of
the requirement that the second-order $m_s$ corrections do not
spoil the non-exotic spectra, and that the $\pi$-nucleon
$\SigmapN$ term lies within the modern phenomenological
range.

There is almost no doubt today that the lightest member of the
exotic antidecuplet has been discovered. We have used its mass and
the latest determinations of the $\pi$-nucleon $\SigmapN$
term~\cite{Sigma} to predict successfully the mass of the
$\Xi_{\overline{10}}$, as also done in~\cite{Prasz} and
\cite{Kopel}. These predictions, however, rely on the
determination of the $\SigmapN$ term which have been varying
over the last 20 years between 45 and 77 MeV. Nevertheless, the
very existence of exotic $\Xi_{\overline{10}}$ and $\Xi^{+}$
states between 1830 and 2000 GeV is an unavoidable prediction of
the chiral soliton models, as can be seen in Fig.
5.\footnote{These numbers do not include the uncertainties of
about 50~MeV due to ${\cal O}(m^2_s)$ corrections.}

Quark models have been modified~\cite{KL,JW} to accommodate
light pentaquarks by introducing quark correlations, otherwise
absent in the naive formulations. The positive parity
of the new exotic states, which is an unproved key prediction of the
soliton models, has been accommodated as well. However, some
versions of the quark models \cite{negpar} as well as lattice
calculations \cite{lattice} and QCD sum rules \cite{QCDSR}
predict negative parity. Therefore,
the measurement of the parities and spins of exotic baryons is one of the
most important experimental challenges. It is, however, a
tall order, especially if one realizes that
the parity and spin of the $\Omega^-$,
whose discovery was a milestone in the foundations
of our present understanding of the strong interactions,
have still not been measured until today \cite{PDG}.

The recent announcement of NA49 of the discovery of some members
of an exotic $\Xi$ multiplet with masses around 1860 MeV would
constitute, if confirmed \cite{Fischer:2004qb}, another success of
the soliton model. As we have already discussed above, the model
is quite flexible in accommodating a $\Xi$ mass in the wide range
between 1830 to 2000 GeV. However, it is encouraging that the mass
reported by NA49 and recent estimate of the $\pi$-nucleon
$\SigmapN$ term~\cite{Sigma} are consistent within the model
accuracy~\cite{Schweitzer:2003fg}. On the other hand, predictions
of the $\Xi_{\overline{10}}$ masses in the correlated CQM lie
below $1800$~MeV~\cite{KL,JW}, possibly indicating the need for
additional degrees of freedom.

One of the most striking predictions of \chiSM\ calculations was
the successful prediction of a narrow decay width for the
$\Theta^+$~\cite{DPP}. Other calculations predicted a larger decay
width~\cite{Weigel},
partly because they lacked the $G_2$ term which is however small,
partly because the model calculations of the remaining $G_{1,2}$
constants gave a smaller cancellation than the phenomenological fit
of~\cite{DPP}, and partly because the larger $\Theta^+$ mass was
used enhancing the phase space factor  $p^3$. We find that the
$\Theta^+$ decay width is suppressed for values of the \chiSM\
couplings that lie close to the ranges favoured in models, and
that it is further suppressed by the SU(3)-breaking effects due to
representation mixing. In comparison, the CQM has available some
suitable dynamical suppression mechanisms based on colour and
spatial overlap arguments \cite{Carlson:2003xb} and selection
rules \cite{Buccella:2004rh}. Another possible suppression
mechanism has been recently proposed within the framework of the
CQM, involving mixing between the two nearly degenerate states
that arise in models with two diquarks and an antiquark
\cite{Karliner:2004qw}.

The narrowness of the $\Theta^+$ in chiral soliton models is far from
being intuitive. It occurs due to the cancellation of the couplings
in the collective decay operator as a conspiracy of the
SU(3) group-theoretical factors and phenomenological values of
these couplings. This cancellation is, however, by no means accidental.
Indeed, in the small soliton limit the cancellation is exact. If in the $\chi$SM
one artificially sets
the soliton size $r_{0}\rightarrow0$, then the model reduces to free
valence quarks which, however, `remember' the soliton structure~\cite{limit}.
In this limit, many quantities are given as ratios
of group-theoretical factors, yielding famous quark model results:
$g_A=5/3$, $\Delta \Sigma =1$ and $\mu_p/\mu_n=-2/3$. Therefore the
small-soliton limit is a very useful theoretical tool for understanding the
predictions of soliton models.

In order to get reliable estimates of the individual couplings, rather
than only of the combinations which enter in the decuplet and
antidecuplet decay widths separately, we have discussed various corrections.
Following~\cite{DPP}, we have multiplied the widths by the appropriate
mass ratios and also by the correction factors due to representation
mixing. These factors are found to be large, so model
predictions for the decay widths suffer from large uncertainties.
Incidentally, these corrections
tend coherently to suppress the width of $\Theta^+$, while the
width of $\Xi_{\overline{10}}$ is coherently enhanced.

Are there any exotics beyond the $\overline{10}$? In soliton models
one gets a tower of exotic states starting with (27, $\frac{3}{2}$),
(35, $\frac{5}{2}$), etc. Whether they can easily be seen is another
issue. As one can see from Fig. 5, the existence of a relatively
light isotriplet of $\Theta_1$
states belonging to the (27, $\frac{3}{2}$) representation,
just a few tens of MeV above the $\Theta^+$,
is quite a robust prediction of the
soliton models. Unlike the antidecuplet $\Theta^+$ though, the decay
widths of (27, $\frac{3}{2}$) states to ordinary octet baryons are
relatively large.
We have estimated $\Gamma_{\Theta^{++}_{27}} \sim 37 \div 66$~MeV,
with a possible enhancement due to the correction factors discussed in
the text. Furthermore, 27 baryons, unlike the $\overline{10}$ ones,
can decay into ordinary decuplet baryons. However, the widths of
these decays are small
and comparable to the decay widths of $\overline{10}$ to 8.
Again in this case the effective decay coupling vanishes
in the small soliton limit discussed above.

Interestingly, another quite robust prediction of the present
model is the existence of the nearly degenerate $I=3/2$ $\Xi$
multiplets in the $\overline{10}$ and (27, $\frac{3}{2}$)
representations. The decay $\Xi(1860)^{-}\rightarrow\Xi^{\ast0}+\pi^{-}$
recently reported by NA49
could be interpreted as an observation of $\Xi_{27}$. However,
all charged states of $\Xi_{27}$ must be found in order to confirm
this hypothesis. Moreover, the rather large total width of the
$\Xi_{27}$ obtained in the present work poses serious challenges for
such an interpretation. On the other hand, in the correlated CQM
such a decay is naturally explained \cite{JW2} as as the decay of the
$\Xi$ isodoublet belonging to a nearly degenerate pentaquark octet.

Therefore, the observation of $\Xi(1860)^{--}$ and $\Xi(1860)^{+}$
decays into decuplet would suggest discovery of yet another tower
of exotic states. However, the non-observation of these decays,
together with positive evidence for $\Xi(1860)^{-}$ and
$\Xi(1860)^{0}$ decays to decuplet baryons, would not rule the
soliton models out immediately. That is because there must be
vibrational excitations \cite{Weigel,Cohen} that we have not
discussed here, among them an octet similar to that predicted by
CQM.

If, however, no other exotics were to be found, how could one get rid of
the whole tower of rotational excitations predicted by the soliton
models? There has been already some discussion in the literature
\cite{Cohen,DP9,Kleb} whether the collective quantization of the rigidly
rotating soliton can be applied to the antidecuplet in the first
place. Surely, the higher the excitations, the more unreliable
is the rigid approximation. Where exactly it breaks down is hard
to say, but it cannot even be excluded that the antidecuplet is the first
and the last exotic representation for which soliton model
predictions still hold.

The confirmed discovery of the $\Theta^+$, together with that of the
$\Xi_{\overline{10}}$\, if it is also confirmed, usher in a new era of hadron
spectroscopy \cite{newexps}.
These developments are already challenging simple versions of the CQM and
\chiSM. Understanding the masses, spin-parities and widths of these
exotic baryons and their undiscovered multiplet partners will require a
new synthesis of methods in non-perturbative QCD, in which elements of
both the CQM and the \chiSM\ may play significant r\^oles.

\acknowledgments

The research of one of us (M.K.) was supported in part by a grant from the
United States-Israel Binational Science Foundation (BSF), Jerusalem.
The present work is supported by the Polish State Committee
for Scientific Research under grant 2 P03B 043 24 (M.P.). This
manuscript has been authored  under Contract No. DE-AC02-98CH10886
with the U. S. Department of Energy.


\begin{thebibliography}{99}

\bibitem{CQM}
J.~J.~Kokkedee, {\it The Quark Model} (Benjamin, New York, 1969);
F.~E.~Close,
arXiv:hep-ph/0311087.

\bibitem{chiSM1}
H.~Weigel, Int.\ J.\ Mod.\ Phys.\ A {\bf  11} (1996) 2419;

\bibitem{chiSM2}
D.~I.~Diakonov, Lectures given at the {\it Advanced Summer School
on Nonperturbative Quantum Field Physics}, Peniscola, Spain, 2-6
June 1997, arXiv:hep-ph/9802298.

\bibitem{Skyrme}
T.~H.~R.~Skyrme,
Proc.\ Roy.\ Soc.\ Lond.\ A {\bf 260} (1961) 127 and Nucl.\ Phys.\ {\bf
31} (1962) 556;
see also
E.~Witten,
Nucl.\ Phys.\ B {\bf 160} (1979) 57 and {\bf 223} (1983) 422, 433;
G.~S.~Adkins, C.~R.~Nappi and E.~Witten,
Nucl.\ Phys.\ B {\bf 228} (1983) 552;
G.~S.~Adkins and C.~R.~Nappi,
Nucl.\ Phys.\ B {\bf 233} (1984) 109.

\bibitem{Guad}
E.~Guadagnini,
Nucl.\ Phys.\ B {\bf 236} (1984) 35.

\bibitem{BEK}
S.J.~Brodsky, J.R.~Ellis and M.~Karliner,
Phys.\ Lett.\ B {\bf 206} (1988) 309.

\bibitem{antidec}
P.~O.~Mazur, M.~A.~Nowak and M.~Prasza{\l}owicz,
Phys.\ Lett.\ B {\bf 147} (1984) 137;
A.~V.~Manohar,
Nucl.\ Phys.\ B {\bf 248} (1984) 19;
M.~Chemtob,
Nucl.\ Phys.\ B {\bf 256} (1985) 600;
S.~Jain and S.~R.~Wadia,
Nucl.\ Phys.\ B {\bf 258} (1985) 713;
M.~P.~Mattis and M.~Karliner,
Phys.\ Rev.\ D {\bf 31} (1985) 2833;
M.~Karliner and M.~P.~Mattis,
Phys.\ Rev.\ D {\bf 34} (1986) 1991;

\bibitem{DPP}
D.~Diakonov, V.~Petrov and M.~V.~Polyakov,
Z.\ Phys.\ A {\bf 359} (1997) 305
[arXiv:hep-ph/9703373].

\bibitem{Weigel}
H.~Weigel,
Eur.\ Phys.\ J.\ A {\bf 2} (1998) 391
[arXiv:hep-ph/9804260].

\bibitem{Prasz}
M.~Prasza{\l}owicz, {\it Proc. of the Workshop on Skyrmions and
Anomalies}, Krak\'ow, 1987, eds. M~Je\.zabek and
M.~Prasza{\l}owicz (World Scientific, Singapore, 1987), p.531;
M.~Prasza{\l}owicz, Phys.\ Lett.\ B {\bf 575} (2003) 234
[arXiv:hep-ph/0308114].

\bibitem{Theta}
T.~Nakano {\it et al.}  [LEPS Collaboration],
Phys.\ Rev.\ Lett.\  {\bf 91} (2003) 012002
[arXiv:hep-ex/0301020].

\bibitem{others}
V.~V.~Barmin {\it et al.}  [DIANA Collaboration],
Phys.\ Atom.\ Nucl.\  {\bf 66} (2003) 1715
[Yad.\ Fiz.\  {\bf 66} (2003) 1763],
hep-ex/0304040;
S.~Stepanyan {\it et al.}  [CLAS Collaboration],
hep-ex/0307018.
%
J.~Barth {\it et al.}  [SAPHIR Collaboration],
hep-ex/0307083;
%
V.~Kubarovsky and S.~Stepanyan  and CLAS Collaboration,
hep-ex/0307088;
%
A.~E.~Asratyan, A.~G.~Dolgolenko and M.~A.~Kubantsev,
hep-ex/0309042.
%
V.~Kubarovsky {\it et al.}, [CLAS Collaboration],
hep-ex/0311046;
A. Airapetian {\it et al.}, [HERMES Collaboration],
arXiv:hep-ex/0312044;
S.~Chekanov, [ZEUS Collaboration],
{\tt http://www.desy.de/f/seminar/Chekanov.pdf};
%
R.~Togoo {\it et al.}, Proc. Mongolian Acad. Sci., {\bf 4} (2003) 2;
A.~Aleev {\it et al.}, [SVD Collaboration],
arXiv:hep-ex/0401024.

\bibitem{BieDo}
L.C Biedenharn and Y. Dothan, {\em Monopolar Harmonics in
SU(3)$_{\rm F}$ as eigenstates of the Skyrme-Witten model for
baryons}, E. Gotsman and G. Tauber (eds.), {\em From SU(3) to
gravity}, p. 15-34.

\bibitem{KL}
M.~Karliner and H.~J.~Lipkin,
Phys.\ Lett.\ B {\bf 595} (2003) 249,
hep-ph/0307243.

\bibitem{JW}
R.~L.~Jaffe and F.~Wilczek,
Phys. Rev. Lett. {\bf 91} (2003) 232003,
[arXiv:hep-ph/0307341].

\bibitem{otherq}
See also
B.~K.~Jennings and K.~Maltman,
arXiv:hep-ph/0308286;
E.~Shuryak and I.~Zahed,
arXiv:hep-ph/0310270.

\bibitem{KB}
P.~Bicudo and G.~M.~Marques,
arXiv:hep-ph/0308073;
D.~E.~Kahana and S.~H.~Kahana,
arXiv:hep-ph/0310026;
F.~J.~Llanes-Estrada, E.~Oset and V.~Mateu,
arXiv:nucl-th/0311020.

\bibitem{Kleb}
N.~Itzhaki, I.~R.~Klebanov, P.~Ouyang and L.~Rastelli,
arXiv:hep-ph/0309305.

\bibitem{Kopel}
H.~Walliser and V.~B.~Kopeliovich, J.\ Exp.\ Theor.\ Phys.\  {\bf
97} (2003) 433 [Zh.\ Eksp.\ Teor.\ Fiz.\  {\bf 124} (2003) 483]
[arXiv:hep-ph/0304058];

\bibitem{Xi}
C.~Alt {\it et al.}  [NA49 Collaboration],
arXiv:hep-ex/0310014.

\bibitem{Fischer:2004qb}
H.~G.~Fischer and S.~Wenig,
arXiv:hep-ex/0401014.

\bibitem{moments}
H.~Walliser, {\it Baryon as Skyrmion soliton}, ed. G.~Holzwarth
(World Scientific, Singapore, 1992), p. 247 and Nucl.\ Phys.\ A
{\bf 548} (1992) 649;
A.~Blotz, D.~Diakonov, K.~Goeke,
N.~W.~Park, V.~Petrov and
P.~V.~Pobylitsa,
Nucl.\ Phys.\ A {\bf 555} (1993) 765.

\bibitem{Sigma}
These numbers are from
M.~M.~Pavan, I.~I.~Strakovsky, R.~L.~Workman and R.~A.~Arndt,
PiN Newslett.\  {\bf 16} (2002) 110
[arXiv:hep-ph/0111066]
For other recent estimates, see:
T.~Inoue, V.~E.~Lyubovitskij, T.~Gutsche and A.~Faessler,
arXiv:hep-ph/0311275 and references therein.

\bibitem{other27}
D.~Borisyuk, M.~Faber and A.~Kobushkin,
arXiv:hep-ph/0307370
and
arXiv:hep-ph/0312213.
%
B.~Wu and B.~Q.~Ma,
arXiv:hep-ph/0312041
and
arXiv:hep-ph/0312326.

\bibitem{Christov}
A.~Blotz, M.~V.~Polyakov and K.~Goeke,
Phys.\ Lett.\ B {\bf 302} (1993) 151;
C.~V.~Christov, A.~Blotz, K.~Goeke, P.~Pobylitsa, V.~Petrov, M.~Wakamatsu
and T.~Watabe,
Phys.\ Lett.\ B {\bf 325} (1994) 467
[arXiv:hep-ph/9312279];
A.~Blotz, M.~Prasza{\l}owicz and K.~Goeke,
Phys.\ Rev.\ D {\bf 53} (1996) 485
[arXiv:hep-ph/9403314].

\bibitem{ThetaWidth}
S. Nussinov,  hep-ph/0307357;
R.~W.~Gothe and S.~Nussinov,
hep-ph/0308230;
%
R.~A.~Arndt, I.~I.~Strakovsky and R.~L.~Workman,
Phys.\ Rev.\ C {\bf 68} (2003) 042201,
nucl-th/0308012 and nucl-th/0311030;
%
J. Haidenbauer and G. Krein,  hep-ph/0309243;
%
R.~N.~Cahn and G.~H.~Trilling,
hep-ph/0311245;
%
A.~Casher and S.~Nussinov,
Phys.\ Lett.\ B {\bf 578} (2004) 124,
hep-ph/0309208.

\bibitem{Cohen}
T.~D.~Cohen,
arXiv:hep-ph/0309111 and
arXiv:hep-ph/0312191;
\hfill\break
%
P.~V.~Pobylitsa,
arXiv:hep-ph/0310221.


\bibitem{DP9}
D.~Diakonov and V.~Petrov,
arXiv:hep-ph/0309203.

\bibitem{PDG}
K.~Hagiwara {\it et al.}  [Particle Data Group Collaboration],
Phys.\ Rev.\ D {\bf 66} (2002) 010001.


\bibitem{DP}
D.~Diakonov and V.~Petrov,
arXiv:hep-ph/0310212.

\bibitem{quadratic}
H.~Yabu and K.~Ando,
Nucl.\ Phys.\ B {\bf 301} (1988) 601.

\bibitem{Leutwyler}
H.~Leutwyler,
Nucl.\ Phys.\ Proc.\ Suppl.\  {\bf 94} (2001) 108
[arXiv:hep-ph/0011049].

\bibitem{Schweitzer:2003fg}
P.~Schweitzer,
arXiv:hep-ph/0312376.

\bibitem{Dorey}
N.~Dorey, J.~Hughes and M.~P.~Mattis,
Phys.\ Rev.\ D {\bf 50} (1994) 5816
[arXiv:hep-ph/9404274].

\bibitem{Yukawa}
%
M.~Uehara,
Prog.\ Theor.\ Phys.\  {\bf 78}, 984 (1987),
%
Prog.\ Theor.\ Phys.\  {\bf 75}, 212 (1986)
[Erratum-ibid.\  {\bf 75}, 464 (1986)];
H.~Verschelde,
Phys.\ Lett.\ B {\bf 215}, 444 (1988).

\bibitem{bijnens}
J.~Bijnens, H.~Sonoda and M.~B.~Wise,
Phys.\ Lett.\ B {\bf 140} (1984) 421.

\bibitem{Prasz11}
M.~Prasza{\l}owicz, Phys. Lett. B {\bf 538} (2004) 96
[arXiv:hep-ph/0311230].

\bibitem{KPG}
H.-Ch. Kim, M. Prasza{\l}owicz and K. Goeke,
 Phys. Rev. D {\bf 61} (2000) 114006 [arXive: hep-ph/9910282].

\bibitem{Ericson}
T.~E.~O.~Ericson, B.~Loiseau and S.~Wycech,
arXiv:hep-ph/0310134.

\bibitem{Bugg:2003bj}
D.~V.~Bugg and M.~D.~Scadron,
arXiv:hep-ph/0312346.

\bibitem{DDPC}
D. Diakonov, private communication.


\bibitem{Jaffearith}
R.~L.~Jaffe,
arXiv:hep-ph/0401187.

\bibitem{Arndt:2003ga}
R.~A.~Arndt, Y.~I.~Azimov, M.~V.~Polyakov, I.~I.~Strakovsky and
R.~L.~Workman,
arXiv:nucl-th/0312126.


\bibitem{MichalPREP}
M. Prasza{\l}owicz, arXiv:hep-ph/0402038.

\bibitem{KadijaJLab}
K. Kadija, talk at
JLab Pentaquark 2003 Workshop, Nov. 2003,
{\tt
www.jlab.org/intralab/calendar/archive03/pentaquark/talks/kadija.pdf}~.

\bibitem{JW2}
R.~Jaffe and F.~Wilczek,
arXiv:hep-ph/0312369.

\bibitem{Bijker}
R.~Bijker, M.~M.~Giannini and E.~Santopinto,
arXiv:hep-ph/0310281.

\bibitem{DC}
J.~J.~Dudek and F.~E.~Close,
arXiv:hep-ph/0311258.

\bibitem{Hyslop:cs}
J.~S.~Hyslop, R.~A.~Arndt, L.~D.~Roper and R.~L.~Workman,
Phys.\ Rev.\ D {\bf 46} (1992) 961;
%
H.~G.~Juengst  [CLAS Collaboration],
arXiv:nucl-ex/0312019.


\bibitem{negpar}
C.~E.~Carlson, C.~D.~Carone, H.~J.~Kwee and V.~Nazaryan,
Phys.\ Lett.\ B {\bf 573} (2003) 101
[arXiv:hep-ph/0307396];
%
R.~Bijker, M.~M.~Giannini and E.~Santopinto,
arXiv:hep-ph/0312380;

\bibitem{lattice}
F.~Csikor, Z.~Fodor, S.~D.~Katz and T.~G.~Kovacs,
JHEP {\bf 0311} (2003) 070
[arXiv:hep-lat/0309090];
%
S.~Sasaki,
arXiv:hep-lat/0310014.

\bibitem{QCDSR}
S.~L.~Zhu,
Phys.\ Rev.\ Lett.\  {\bf 91} (2003) 232002
[arXiv:hep-ph/0307345];
%
J.~Sugiyama, T.~Doi and M.~Oka,
arXiv:hep-ph/0309271.

\bibitem{Carlson:2003xb}
C.~E.~Carlson, C.~D.~Carone, H.~J.~Kwee and V.~Nazaryan,
arXiv:hep-ph/0312325.

\bibitem{Buccella:2004rh}
F.~Buccella and P.~Sorba,
arXiv:hep-ph/0401083.

\bibitem{Karliner:2004qw}
M.~Karliner and H.~J.~Lipkin,
arXiv:hep-ph/0401072.

\bibitem {limit}M. Prasza\l owicz, A. Blotz and K. Goeke, Phys. Lett.
\textbf{B354} (1995) 415 [hep-ph/9505328];
M. Prasza\l owicz, T. Watabe and K. Goeke, Nucl. Phys.
\textbf{A647} (1999) 49 [hep-ph/9806431].

\bibitem{newexps}
A.~W.~Thomas, K.~Hicks and A.~Hosaka,
arXiv:hep-ph/0312083.
%
C.~Hanhart {\it et al.},
arXiv:hep-ph/0312236.
%
S.~Armstrong, B.~Mellado and S.~L.~Wu,
arXiv:hep-ph/0312344.
%
M.~Bleicher, F.~M.~Liu, J.~Aichelin, T.~Pierog and K.~Werner,
arXiv:hep-ph/0401049.
%
T.E.~Browder, I.R.~Klebanov and D.R.~Marlow,
arXiv:hep-ph/0401115;

\end{thebibliography}
\end{document}